\documentclass[10pt,journal,compsoc]{IEEEtran}

\ifCLASSOPTIONcompsoc
  \usepackage[nocompress]{cite}
\else
  \usepackage{cite}
\fi
\usepackage[skins]{tcolorbox}
\usepackage{tabularx}
\usepackage{caption}
\usepackage{adjustbox}
\usepackage{threeparttable}
\usepackage{subcaption}
\usepackage{balance}
\usepackage[]{enumitem}
\usepackage{makecell}
\usepackage{amsmath}
\usepackage{booktabs}
\usepackage{url}
\usepackage[hidelinks]{hyperref}
\usepackage{ragged2e}
\usepackage{csquotes}
\usepackage[T1]{fontenc}
\usepackage{tgcursor}
\usepackage{threeparttable}
\usepackage{multirow}

\renewenvironment{quote}
  {\list{}{\rightmargin=15pt \leftmargin=15pt}%
   \item\relax}
  {\endlist}

\DeclareMathOperator*{\mean}{mean} 

\newcommand{\Code}[1]{\texttt{#1}}
\newcommand*{\MyIndent}{\hspace*{0.3cm}}

\newcommand{\RQ}[2]{\textbf{RQ#1:} \textit{#2}}

\tcbset{
  my box/.style={
    enhanced,
    colframe=#1!80,
    colback=#1!10,
    attach boxed title to top left={xshift=0.2cm, yshift=-0.2cm},
    boxed title style={
      colback=#1!80,
      outer arc=0pt,
      arc=0pt,
      top=0pt,
      bottom=0pt,
    },
  },
}
\newtcolorbox{result-rq}[1]{
  my box=black,
  title=#1,
  boxrule=1.2pt,top=6pt,bottom=3.5pt,left=6pt,right=6pt
}

\begin{document}

\title{Automating Dependency Updates in Practice: An Exploratory Study on GitHub Dependabot}

\author{Runzhi He, Hao He, Yuxia Zhang, Minghui Zhou
\IEEEcompsocitemizethanks{\IEEEcompsocthanksitem Runzhi He, Hao He, and Minghui Zhou are with School of Computer Science, Peking University, Beijing, China, and Key Laboratory of High Confidence Software Technologies, Ministry of Education, Beijing, China. Runzhi He and Hao He contributed equally to this work. Minghui Zhou is the corresponding author.\\
Email: \href{mailto:rzhe@pku.edu.cn}{rzhe@pku.edu.cn}, \href{mailto:heh@pku.edu.cn}{heh@pku.edu.cn}, \href{mailto:zhmh@pku.edu.cn}{zhmh@pku.edu.cn}
\IEEEcompsocthanksitem Yuxia Zhang is with School of Computer Science and Technology, Beijing Institute of Technology, Beijing, China.\\
Email: \href{mailto:yuxiazh@bit.edu.cn}{yuxiazh@bit.edu.cn}}
\thanks{Manuscript received July 31, 2022; Revised March 16, 2023}}

\markboth{IEEE Transactions on Software Engineering}%
{He \MakeLowercase{\textit{et al.}}: Automating Dependency Updates in Practice: An Exploratory Study on GitHub Dependabot}

\IEEEtitleabstractindextext{
\begin{abstract}
Dependency management bots automatically open pull requests to update software dependencies on behalf of developers. Early research shows that developers are suspicious of updates performed by dependency management bots and feel tired of overwhelming notifications from these bots. Despite this, dependency management bots are becoming increasingly popular. Such contrast motivates us to investigate Dependabot, currently the most visible bot on GitHub, to reveal the effectiveness and limitations of state-of-art dependency management bots. We use exploratory data analysis and a developer survey to evaluate the effectiveness of Dependabot in keeping dependencies up-to-date, interacting with developers, reducing update suspicion, and reducing notification fatigue. We obtain mixed findings. On the positive side, projects do reduce technical lag after Dependabot adoption and developers are highly receptive to its pull requests. On the negative side, its compatibility scores are too scarce to be effective in reducing update suspicion; developers tend to configure Dependabot toward reducing the number of notifications; and 11.3\% of projects have deprecated Dependabot in favor of other alternatives. The survey confirms our findings and provides insights into the key missing features of Dependabot. Based on our findings, we derive and summarize the key characteristics of an ideal dependency management bot which can be grouped into four dimensions: configurability, autonomy, transparency, and self-adaptability.
\end{abstract}
}

\IEEEdisplaynontitleabstractindextext
\IEEEpeerreviewmaketitle

\maketitle

\section{Introduction}
\label{sec:introduction}

To update or not to update, that is the question haunting software engineers for decades.
The software engineering ``gurus'' would argue that keeping software dependencies\footnote{
In this paper, we keep inline with developers' common terminologies and use the term ``dependency'' to refer to any software that other software \textit{depends} on, such as libraries, packages, frameworks, development tools, etc.
Dependencies are often managed by a dependency management tool (e.g., npm for JavaScript) and declared in a dependency specification file (e.g., \texttt{package.json} for npm projects).
}
up-to-date minimizes technical debt, increases supply chain security, and ensures software project sustainability in the long term~\cite{
winters2020software}.
Nonetheless, it requires not only substantial effort but also extra responsibility from developers.
Consequently, 
many developers adhere to the practice of ``if it ain't broke, don't fix it'' and the majority of existing software systems use outdated dependencies~\cite{DBLP:journals/ese/KulaGOII18}. 

One promising solution for this dilemma is to use bots to automate all dependency updates.
Therefore, \textit{dependency management bots} are invented to automatically open pull requests (PRs) to update dependencies in a collaborative coding platform (e.g., GitHub) in the hope of saving developer effort.
Recently, dependency management bots are increasingly visible and gaining high momentum among practitioners.
The exemplars of these bots, including Dependabot~\cite{Dependabot}, Renovate Bot~\cite{RenovateBot}, PyUp~\cite{PyUp}, and Synk Bot~\cite{SynkBot}, have opened millions of PRs on GitHub~\cite{DBLP:conf/botse-ws/WyrichGHM21} and are adopted by a variety of industry teams (according to their websites).

However, the simple idea of using a bot does not save the world.
The early work of Mirhosseini and Parnin~\cite{DBLP:conf/kbse/MirhosseiniP17} on Greenkeeper~\cite{Greenkeeper} reveals that: 
only 32\% of Greenkeeper PRs are merged because developers are suspicious of whether a bot PR will break their code 
(i.e., \textit{update suspicion}) and feel annoyed about a large number of bot PRs (i.e., \textit{notification fatigue}).
Since then, similar bots have emerged, evolved, and gained high popularity, among them the most visible one on GitHub is Dependabot~\cite{DBLP:conf/botse-ws/WyrichGHM21} with many improvements (Section~\ref{sec:dependabot}). 
However, it remains unknown to what extent can these bots overcome the two limitations of Greenkeeper 
identified by Mirhosseini and Parnin~\cite{DBLP:conf/kbse/MirhosseiniP17} in 2017.

To shed light on improving dependency management bots and software engineering bots in general, we present an exploratory study on Dependabot.
Our study answers the following four research questions (RQs) to empirically evaluate the effectiveness of Dependabot version update in different dimensions (detailed motivations in Section~\ref{sec:rqs}): 
\begin{itemize}[leftmargin=15pt]
\RaggedRight
    \item \RQ{1}{To what extent does Dependabot reduce the technical lag of a project after its adoption?}
    \item \RQ{2}{How actively do developers respond to and merge pull requests opened by Dependabot?}
    \item \RQ{3}{How effective is Dependabot's compatibility score in allaying developers' update suspicion?}
    \item \RQ{4}{How do projects configure Dependabot for automating dependency updates?}
\end{itemize}
As we find that many projects have deprecated Dependabot in favor of other alternatives, we ask an additional RQ:
\begin{itemize}[leftmargin=15pt]
\RaggedRight
    \item \RQ{5}{How do projects deprecate Dependabot and what are the developers' desired features for Dependabot?}
\end{itemize}

To answer the RQs, we sample 1,823 popular and actively maintained GitHub projects as the study subjects. 
We conduct exploratory data analysis on 502,752 Dependabot PRs from these projects and use a survey of 131 developers to triangulate our findings.
Our findings provide empirical characterizations of Dependabot's effectiveness in various dimensions.
More importantly, we discover important limitations of Dependabot (a state-of-the-art bot) in overcoming update suspicion and notification fatigue, along with the missing features for overcoming the limitations.
Based on the findings, we summarize four key properties of an ideal dependency management bot (i.e., configurability, autonomy, transparency, and self-adaptability) as a roadmap for software engineering researchers and bot designers.

\section{Background and Related Work}
\label{sec:background}

\subsection{Dependency Update}

In modern software development, updating dependencies is not only important but also non-trivial.
A typical software project may have tens to thousands of dependencies and each of the outdated ones induces risks~\cite{SoftwareSupplyChain}.
However, each update may contain breaking changes which can be hard to discover and fix~\cite{DBLP:journals/tosem/BogartKHT21}.
This situation inspires research into understanding update practices, designing metrics, and inventing approaches to support dependency updates.

Bavota et al.~\cite{DBLP:journals/ese/BavotaCPOP15} find that updates in the Apache ecosystems are triggered by major changes or a large number of bug fixes, but may be prevented by API removals.
Kula et al.~\cite{DBLP:journals/ese/KulaGOII18} discover that 81.5\% of the 4600 studied Java/Maven projects on GitHub still keep outdated dependencies due to lack of awareness and extra workload.
Pashchenko et al.~\cite{DBLP:conf/ccs/PashchenkoVM20} find through semi-structured interviews that developers face trade-offs when updating dependencies (e.g., vulnerabilities, breaking changes, policies).

Researchers have proposed measurements to quantify the ``freshness'' or ``outdatedness'' of software dependencies and applied them to various software ecosystems. 
Cox et al.~\cite{DBLP:conf/icse/CoxBEV15} propose several metrics to quantify ``dependency freshness'' and evaluate them on a dataset of industrial Java systems.
A series of studies~\cite{DBLP:conf/oss/Gonzalez-Barahona17, DBLP:conf/icsm/DecanMC18, DBLP:conf/icsr/ZeroualiCMRG18, 
DBLP:journals/smr/ZeroualiMGDCR19, DBLP:conf/apsec/StringerTB020, DBLP:journals/ese/ZeroualiMDGR21} introduce the notion of \textit{technical lag}, a metric for measuring the extent of project dependencies lagging behind their latest releases, and investigate the evolution of technical lag in Debian~\cite{DBLP:conf/oss/Gonzalez-Barahona17}, npm~\cite{DBLP:conf/icsm/DecanMC18, DBLP:conf/icsr/ZeroualiCMRG18, DBLP:journals/smr/ZeroualiMGDCR19}, the Libraries.io dataset~\cite{DBLP:conf/apsec/StringerTB020}, and Docker images~\cite{DBLP:journals/ese/ZeroualiMDGR21}.
They find that technical lag tends to increase over time, induces security risks, and can be mitigated using semantic versioning. 

There has been a long line of research in software engineering for supporting the automated update of software.
Since API breaking changes form the majority of update cost, most studies propose automated approaches to match and adapt evolving APIs (e.g., \cite{DBLP:conf/icsm/ChowN96, DBLP:conf/icse/HenkelD05, DBLP:journals/tse/XingS07,  DBLP:conf/oopsla/NguyenNWNKN10, DBLP:journals/tosem/DagenaisR11}).
However, Cossette and Walker~\cite{DBLP:conf/sigsoft/CossetteW12} reveal through manual analysis that real API adaptation tasks are complex and beyond the capability of previous automated approaches.
Recently, research interest in automated API adaptation is surging again with works on Java~\cite{
DBLP:conf/kbse/Huang0PW021}, JavaScript~\cite{DBLP:conf/icse/NielsenTM21}, Python~\cite{DBLP:conf/icsm/HaryonoT0LJ21a}, Android~\cite{
DBLP:journals/ese/HaryonoTLJLKSM22}, etc.

On the other hand, practitioners often take conservative update approaches: upstream developers typically use semantic versioning~\cite{SemVer} for signaling version compatibility; downstream developers perform most updates manually and detect incompatibilities through release notes, compilation failures, and regression testing.
Unfortunately, studies~\cite{DBLP:conf/issta/MostafaRW17, DBLP:journals/jss/RaemaekersDV17, DBLP:journals/jss/HejderupG22,2022ICPC-ReleaseNote} reveal that none of them work well in guaranteeing update compatibility.
Generally, providing such guarantees is still a challenging open problem~\cite{DBLP:conf/oopsla/LamDP20}.

\subsection{Dependency Management Bots}

Perhaps the most noticeable automation effort among practitioners is dependency management bots. 
These bots automatically create pull requests (PRs) to update dependencies either immediately after a new release is available or when a security vulnerability is discovered in the currently used version.
In other words, dependency management bots solve the lack of awareness problem~\cite{DBLP:journals/ese/KulaGOII18} by automatically pushing update notifications to developers. 

Mirhosseini and Parnin~\cite{DBLP:conf/kbse/MirhosseiniP17} conduct a pioneering study on Greenkeeper and find that developers update dependencies 1.6x more frequently with Greenkeeper, but only 32\% of Greenkeeper PRs are merged due to two major limitations:

\begin{itemize}[leftmargin=15pt]
    \item \textbf{Update Suspicion:} If an automated update PR breaks their code, developers immediately become suspicious of subsequent PRs and are reluctant to merge them.
    \item \textbf{Notification Fatigue:} If too many automated update PRs are generated, developers may feel annoyed about the notifications and simply ignore all the update PRs.
\end{itemize}
Rombaut et al.~\cite{rombaut2022there} find that Greenkeeper issues for in-range breaking updates induce a large maintenance overhead, and many of them are false alarms caused by project CI issues.

The limitations of Greenkeeper align well with the challenges revealed in the software engineering (SE) bot literature.
Wessel et al.~\cite{DBLP:journals/pacmhci/WesselSSWPCG18} find that SE bots on GitHub have interaction problems and provide poor decision-making support.
Erlenhov et al.~\cite{DBLP:conf/sigsoft/ErlenhovN020} 
identify two major challenges in ``Alex''
bot (i.e., SE bots that autonomously perform simple tasks) design: establishing trust and reducing interruption/noise.
Wyrich et al.~\cite{DBLP:conf/botse-ws/WyrichGHM21} find that bot PRs have a lower merge rate and need more time to be interacted with and merged.
Two subsequent studies by Wessel et al.~\cite{DBLP:journals/pacmhci/WesselWSG21, DBLP:conf/icse/WesselAWCSGS22} qualitatively show that noise is the central challenge in SE bot design but it can be mitigated by certain design strategies and the use of a ``meta-bot.''
Shihab et al.~\cite{shihab2022present} draw a picture of SE bot technical and socio-economicd challenges. 
Santhanam et al.~\cite{DBLP:journals/peerj-cs/SanthanamHSW22} provide a systematic mapping of the SE bot literature.

Since Mirhosseini and Parnin~\cite{DBLP:conf/kbse/MirhosseiniP17}, many other bots have emerged for automating dependency updates, such as Dependabot~\cite{Dependabot} (preview release in May 2017) and Renovate Bot~\cite{RenovateBot} (first release in January 2017).
Greenkeeper itself reaches end-of-life in June 2020 and its team merged with Synk Bot~\cite{SynkBot}.
All these bots are widely used: according to Wyrich et al.~\cite{DBLP:conf/botse-ws/WyrichGHM21}, they opened the vast majority of bot PRs on GitHub (six out of the top seven). 
The top two are occupied by Dependabot~\cite{Dependabot} and Dependabot Preview~\cite{DependabotPreview} with $\sim$3 million PRs and $\sim$1.2 million PRs, respectively.
Erlenhov et al.~\cite{DBLP:journals/peerj-cs/ErlenhovNL22} find that under a strict SE bot definition, almost all bots in an existing bot commit dataset~\cite{DBLP:conf/msr/DeyMPFVFM20} are dependency management bots and they are frequently adopted, discarded, switched, and even simultaneously used by GitHub projects, indicating a fierce competition among them.


\subsection{Dependabot}
\label{sec:dependabot}

Among different dependency management bots, Dependabot~\cite{Dependabot} is the most visible one in GitHub projects~\cite{DBLP:conf/botse-ws/WyrichGHM21}. 
Dependabot Preview was launched in 2017~\cite{DependabotInterview} and acquired by GitHub in 2019~\cite{DependabotAquisition}.
In August 2021, it was shut down in favor of the new, GitHub native Dependabot~\cite{DependabotShutdown} operating since June 2020, which offers two main services:
\begin{itemize}[leftmargin=15pt]
    \item \textbf{Dependabot version update~\cite{DependabotConfig}:} If a configuration file named \Code{dependabot.yml} is added to a GitHub repository, Dependabot will begin to open PRs that update project dependencies to the latest version. Developers can specify the exact Dependabot behavior in \Code{dependabot.yml} (e.g., update interval and the max number of PRs). 
    \item \textbf{Dependabot security update~\cite{DependabotAlertAccess}:} Dependabot scans the entire GitHub to find repositories with vulnerable dependencies. Even if no \Code{dependabot.yml} is supplied, Dependabot still alerts repository owners and repository owners can tell Dependabot to open PRs that update vulnerable dependencies to their patched versions.
\end{itemize}

\begin{figure}[!t]
\centering
\includegraphics[width=0.78\linewidth]{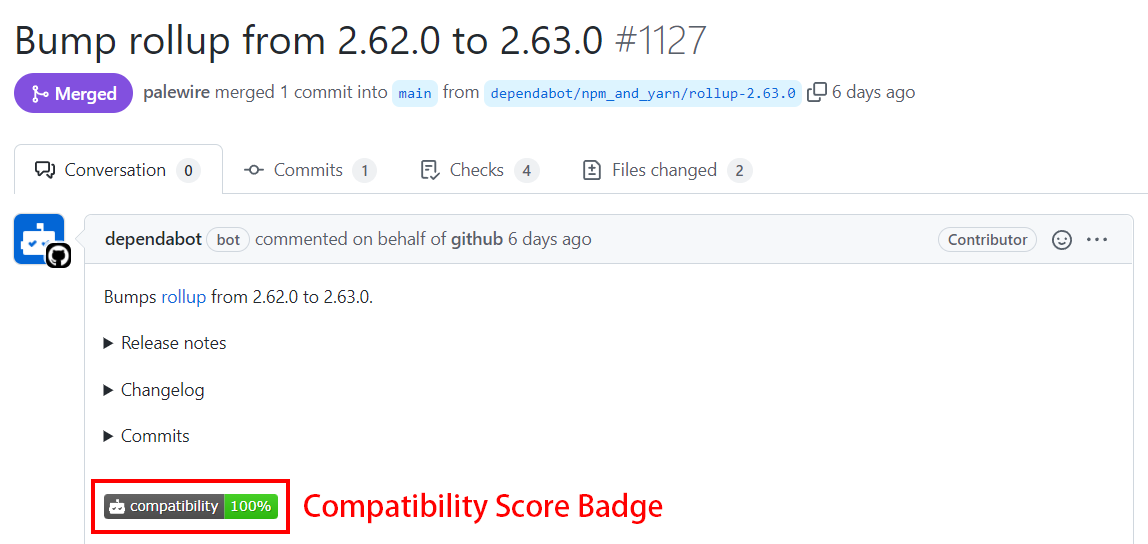}
\vspace{-1mm}
\caption{A pull request opened by Dependabot}
\label{fig:depbotpr}
\vspace{-4mm}
\end{figure}

Figure~\ref{fig:depbotpr} shows an example Dependabot PR~\cite{datadesk/baker/pull/1127}.
Apart from all other details, one especially interesting Dependabot feature is the \textbf{compatibility score} badge. 
According to GitHub documentation~\cite{DependabotCompatScore}:
\textit{An update's compatibility score is the percentage of CI runs that passed when updating between specific versions of the dependency.}
In other words, the score uses the large-scale regression testing data available in GitHub CI test results to estimate the risk of breaking changes in a dependency update.
This looks like a promising direction for solving the \textbf{update suspicion} problem, as previous studies have shown that project test suites are often unreliable in detecting update incompatibilities~\cite{DBLP:journals/jss/HejderupG22} and the false alarms introduce significant maintenance overhead~\cite{rombaut2022there}.
However, the score's effectiveness in practice remains unknown.

For the \textbf{notification fatigue} problem, Wessel et al.~\cite{DBLP:journals/pacmhci/WesselWSG21} suggest SE bots offer flexible configurations and send only relevant notifications.
Both solutions have been (principally) implemented by Dependabot, but it is still unclear whether the specific configuration options and notification strategies taken by Dependabot are really effective in practice.
Alfadel et al.~\cite{DBLP:conf/msr/AlfadelCSM21} find that developers receive Dependabot security PRs well: 65.42\% of PRs are merged and most are merged within a day.
However, security PRs only constitute a small portion of Dependabot PRs (6.9\% in our dataset), and developers perceive security updates as highly relevant~\cite{DBLP:conf/ccs/PashchenkoVM20}.
The effectiveness of Dependabot version update in general seems to be more problematic.
Soto-Valero et al.~\cite{DBLP:conf/sigsoft/Soto-ValeroDB21} find that Dependabot opens many PRs on bloated dependencies.
Cogo and Hassan~\cite{cogo2022understanding} provides evidence on how the configuration of Dependabot causes issues for developers.
As stated by two developers in a GitHub issue~\cite{caddyserver/caddy/pull/4317}: 
1) \textit{I think we'd rather manage dependency upgrades ourselves, on our own time. We've been frequently bitten by dependency upgrades causing breakages. We tend to only upgrade dependencies when we're close to being ready to cut a release.}
2) \textit{Also Dependabot tends to be pretty spammy, which is rather annoying.}

To the best of our knowledge, a comprehensive empirical investigation into the adoption of the Dependabot version update service is still lacking.
Such knowledge from Dependabot can help the formulation of general design guidelines for dependency management bots and unveil important open challenges for fulfilling these guidelines.

\section{Research Questions}
\label{sec:rqs}

Our study goal is to evaluate the practical effectiveness of the \textbf{Dependabot version update} service.
In this Section, we elaborate on the motivation of each RQ toward this goal. 

The Dependabot version update service is designed to make developers aware of new versions and help them keep project dependencies up-to-date. 
To quantitatively evaluate the extent to which Dependabot fulfills its main design purpose
(i.e., \textit{keeping dependencies up-to-date}),
we reuse metrics from the technical lag literature~\cite{DBLP:conf/icsm/DecanMC18,DBLP:journals/smr/ZeroualiMGDCR19} and ask:

\begin{quote}
  \RQ{1}{To what extent does Dependabot reduce the technical lag of a project after its adoption?}
\end{quote}

To help developers keep dependencies up-to-date, Dependabot intervenes by automatically creating update PRs when new versions become available, after which developers can interact with (e.g., comment, merge) these PRs.
We evaluate the effectiveness of this interaction process by measuring the extent to which \textit{developers interact smoothly with Dependabot PRs}, forming the next RQ:


\begin{quote}
   \RQ{2}{How actively do developers respond to and merge pull requests opened by Dependabot?} 
\end{quote}



One major limitation of Greenkeeper is that developers tend to be suspicious of whether a dependency update will introduce break changes~\cite{DBLP:conf/kbse/MirhosseiniP17} (i.e., \textit{update suspicion}).
On the other hand, Dependabot helps developers establish confidence on update PRs using the \textit{compatibility score} feature (Section~\ref{sec:dependabot}).
To quantitatively evaluate the effectiveness of this feature against update suspicion, we ask:

\begin{quote}
  \RQ{3}{How effective is Dependabot's compatibility score in allaying developers' update suspicion?}
\end{quote}

The other major limitation of Greenkeeper is that developers tend to be overwhelmed by a large number of update PRs~\cite{DBLP:conf/kbse/MirhosseiniP17} (i.e., \textit{notification fatigue}).
On the other hand, Dependabot provides flexible configuration options for controlling the amount notifications (Section~\ref{sec:dependabot}). 
To explore how developers configure (and re-configure) the number of notifications generated by Dependabot, we study real-world Dependabot configurations and ask:

\begin{quote}
  \RQ{4}{How do projects configure Dependabot for automating dependency updates?}
\end{quote}


During our analysis, we discover that a non-negligible portion of projects in our studied corpus have deprecated Dependabot and migrated to other alternatives.
As an in-depth retrospective analysis of the reasons behind these deprecations can help reveal important Dependabot limitations and future improvement directions, we ask:

\begin{quote}
  \RQ{5}{How do projects deprecate Dependabot and what are the developers' desired features for Dependabot?}
\end{quote}



\section{Study Design}
\label{sec:datacollection}

\begin{figure}[t]
  \centering
  \includegraphics[width=\linewidth]{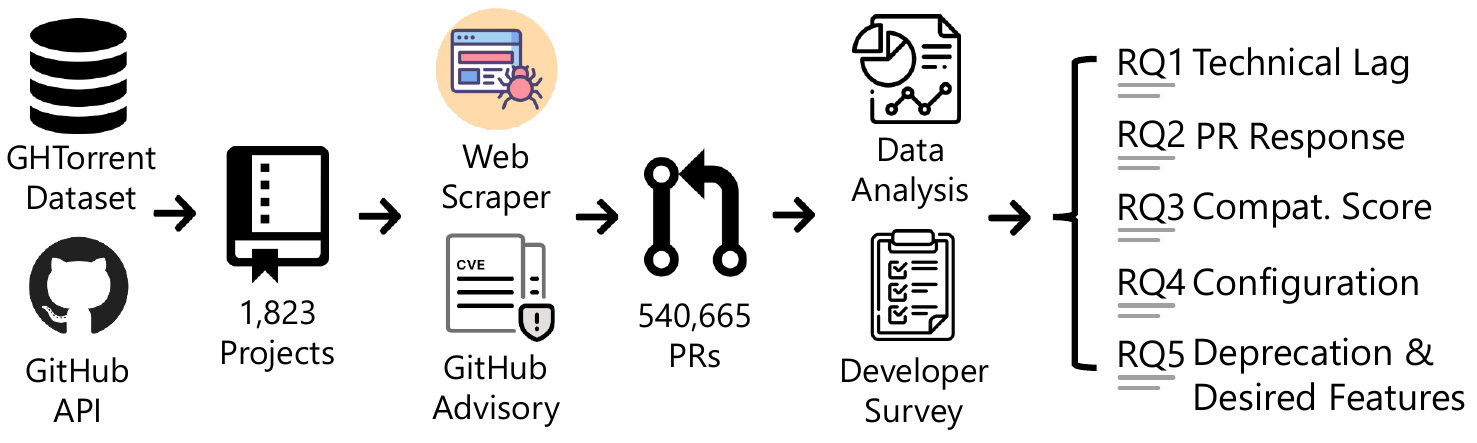}
\vspace{-4mm}
  \caption{An overview of our study}
  \label{fig:workflow}
\vspace{-1mm}
\end{figure}

An overview of our study is shown in Figure~\ref{fig:workflow}.
The study follows a mix-method study design where we obtain results from repository data analysis and triangulate them with a developer survey.
In this Section, we introduce the data collection and survey methods.
The specific analysis methods will be presented along with their results in Section \ref{sec:results}.

\subsection{Data Collection} 

\textbf{Project Selection.}
As the first step, we need to collect a sample of engineered and maintained GitHub projects using or once used Dependabot version update in their workflow.
We focus on the GitHub native Dependabot (released on June 1, 2020) and do not include Dependabot Preview in our study because the former provides much richer features and allows us to obtain the latest, state-of-the-art results. 

We begin with the latest dump of GHTorrent~\cite{Gousi13} (released on March 6, 2021), a large-scale dataset of GitHub projects widely used in software engineering research (e.g., \cite{DBLP:conf/botse-ws/WyrichGHM21, DBLP:journals/jss/HejderupG22}).
We find a noticeable gap in the GHTorrent dataset from July 2019 to early January 2020 (also observed by Wyrich et al.~\cite{DBLP:conf/botse-ws/WyrichGHM21}).
Focusing solely on GitHub native Dependabot allows us to circumvent threats caused by this gap because all its PRs are created after January 2020.

We select projects with at least 10 merged Dependabot PRs to keep only projects that have used Dependabot to some degree.\footnote{The threshold is chosen intuitively as knowing the global population (i.e., all GitHub projects that have adopted Dependabot version update and merged Dependabot PRs) would require scanning \texttt{dependabot.yml} files which is not feasible using GHTorrent.}
To filter out irrelevant, low-quality, or unpopular projects, we retain only non-fork projects with at least 10 stars, as inspired by previous works~\cite{DBLP:journals/ese/MunaiahKCN17, DBLP:conf/sigsoft/HeHGZ21, DBLP:conf/sigsoft/Soto-ValeroDB21}.\footnote{
Munaiah et al.~\cite{DBLP:journals/ese/MunaiahKCN17} show that a simple threshold of 10 stars can already reach a precision of 97\% when classifying engineered GitHub projects.
Other filtering steps also contribute to the overall quality of our final project sample (e.g., sustained activity).
}
Since projects without sustained activities may not perform dependency updates on a regular basis and induce noise in technical lag analysis (\textbf{RQ1}), we query GitHub APIs~\cite{GitHubAPI} and retain projects with a median weekly commit of at least one in the past year.
To exclude projects that have never utilized Dependabot version update, we clone and retain only projects with some \Code{git} change history on \Code{dependabot.yml}.
After all the filtering steps, we end up with 1,823 projects.

\begin{table}[b]
  \scriptsize
  \centering
  \vspace{-2mm}
  \caption{Statistics of the Projects and Survey Respondents}
  \vspace{-2mm}
  \label{tab:projectstats}
  \begin{tabular}
  {clrrc}
  \toprule
    & Statistics & Mean & Median & Distribution \\
    \midrule
    \parbox[t]{-2mm}{\multirow{10}{*}{\rotatebox[origin=c]{90}{\textbf{Projects}}}} 
    & \# of Stars & 1423.92 & 66.00 & \adjustimage{height=0.47cm,valign=m}{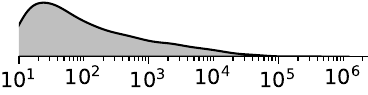}\\
    & \# of Commits & 2837.11 & 1040.50 &  \adjustimage{height=0.47cm,valign=m}{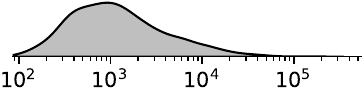}\\
    & \# of Contributors & 26.50 & 12.00 & \adjustimage{height=0.47cm,valign=m}{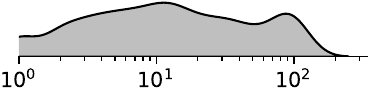}\\
    & Lines of Code (thousands) & 98.18 & 19.89 & \adjustimage{height=0.47cm,valign=m}{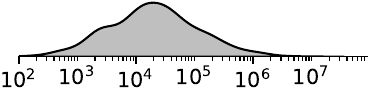}\\
    & \# of Commits per Week & 10.07 & 4.00 &  \adjustimage{height=0.47cm,valign=m}{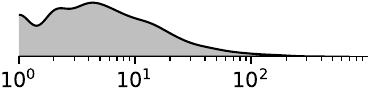}\\
    & Age at Adoption (days) & 1018.18 & 714.00 & \adjustimage{height=0.47cm,valign=m}{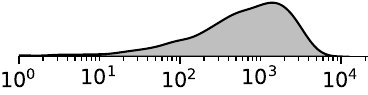}\\
    & \# of Dependabot PRs & 304.56 & 204.00 & \adjustimage{height=0.47cm,valign=m}{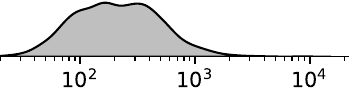}\\
    \midrule
    \parbox[t]{-2mm}{\multirow{6}{*}{\rotatebox[origin=c]{90}{\scriptsize \textbf{Respondents}}}} 
    & \# of Dependabot Interactions & 644.54 & 410.00 & \adjustimage{height=0.47cm,valign=m}{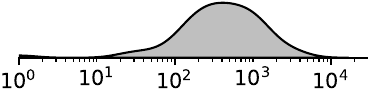}\\
    & \# of Commits & 477.00 & 331.50 & \adjustimage{height=0.47cm,valign=m}{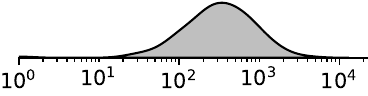}\\
    & \# of Followers & 168.00 &  53.50 & \adjustimage{height=0.47cm,valign=m}{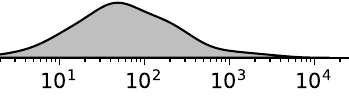}\\
    & Years of Experience (GitHub) & 10.37 & 10.68 & \adjustimage{height=0.47cm,valign=m}{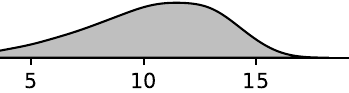}\\
  \bottomrule
  \end{tabular}
\end{table}

\begin{table*}
  \scriptsize
  \caption{Survey Questions and Their Results (131 Responses in Total)}
\vspace{-2mm}
  \label{tab:survey}
  \begin{threeparttable}
  \begin{tabularx}{\textwidth}{X|c|c}
  \toprule
    \textbf{5-Point Likert-Scale Questions} & Distribution & Avg. \\
    \midrule
    (\textbf{RQ1}) Dependabot helps my project keep all dependencies up-to-date. & \adjustimage{height=1.1em,valign=m}{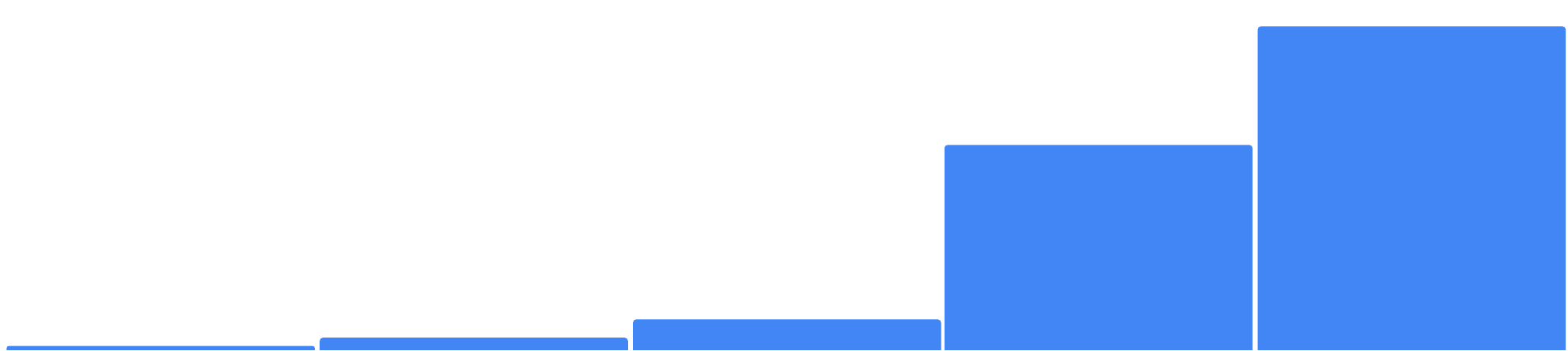} & 4.44 \\
    \midrule
    (\textbf{RQ2}) Dependabot PRs do not require much work to review and merge. & 
    \adjustimage{height=1.1em,valign=m}{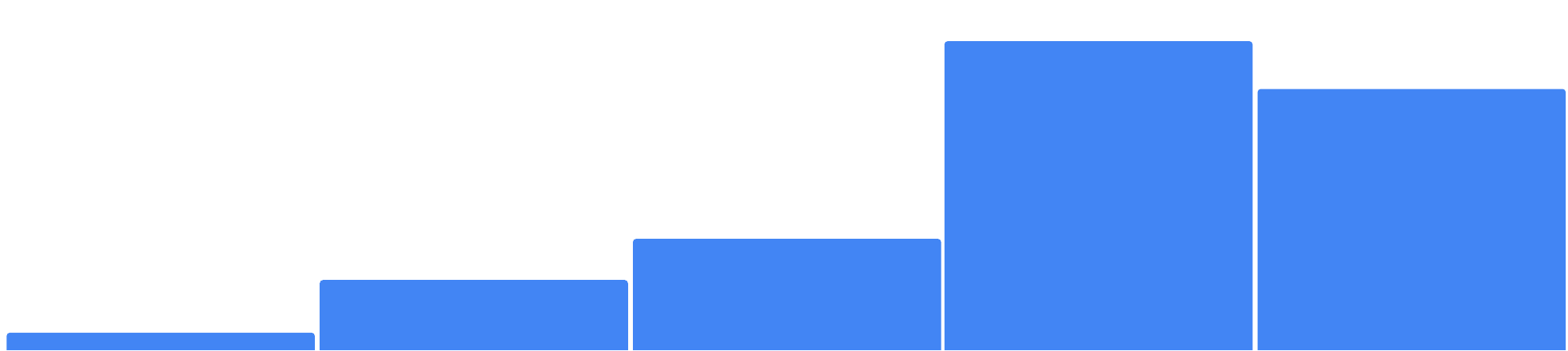} & 3.94 \\
    (\textbf{RQ2}) I respond to a Dependabot PRs fast if it can be safely merged. & 
    \adjustimage{height=1.1em,valign=m}{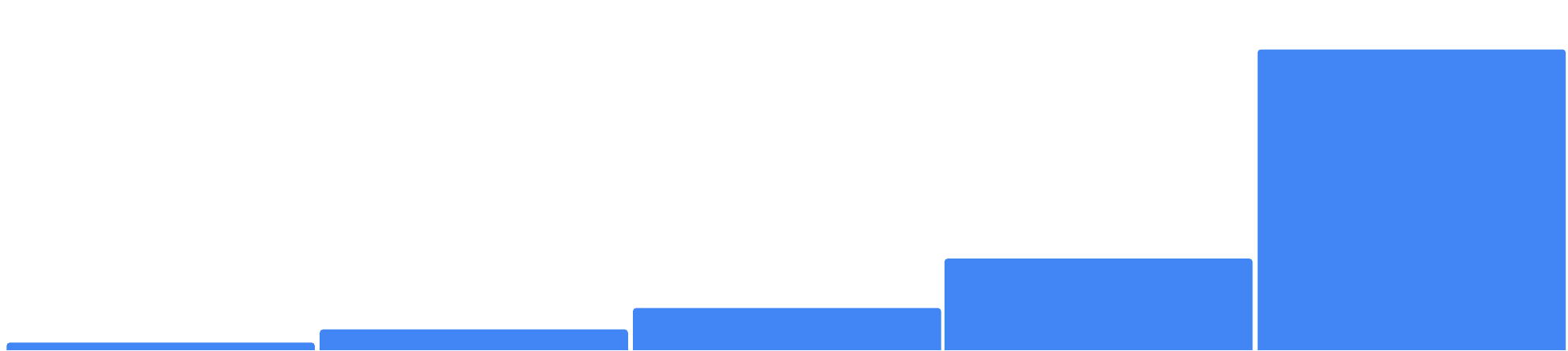} & 4.42 \\
    (\textbf{RQ2}) I ignore the Dependabot PR or respond slower if it cannot be safely merged. &  \adjustimage{height=1.1em,valign=m}{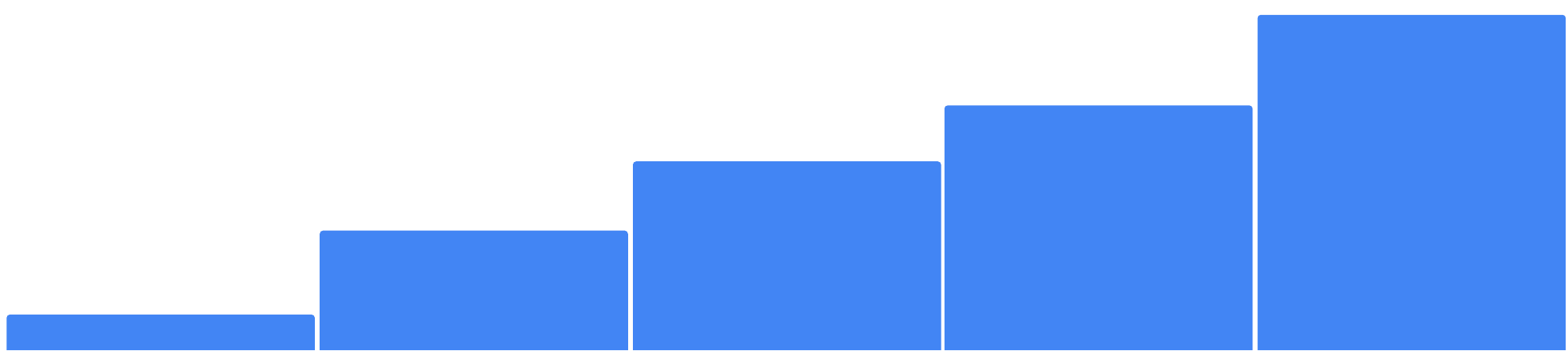} & 3.78 \\
    (\textbf{RQ2}) I handle a Dependabot PR with higher priority if it updates a vulnerable dependency. &  \adjustimage{height=1.1em,valign=m}{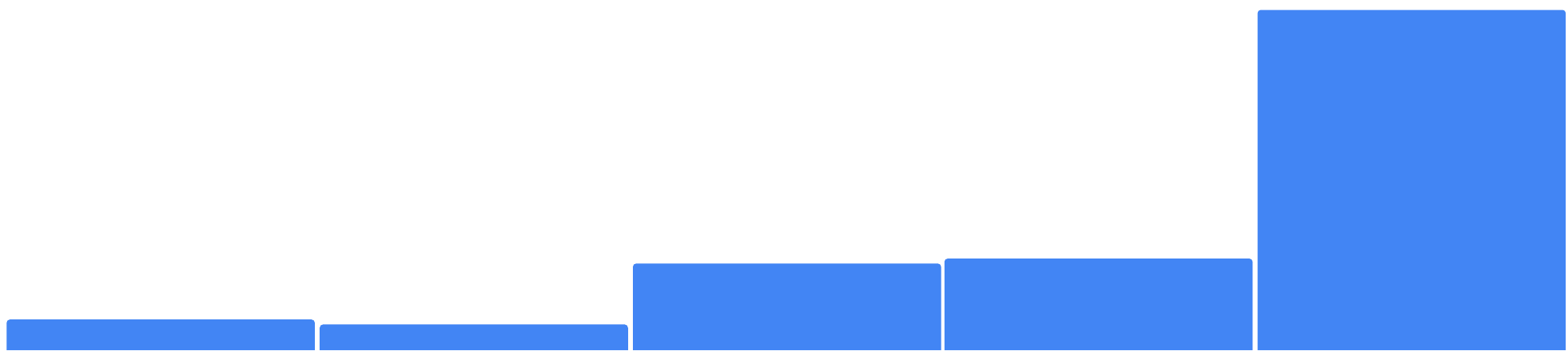} & 4.19 \\
    (\textbf{RQ2}) It requires more work to review and merge a Dependabot PR if it updates a vulnerable dependency. &  \adjustimage{height=1.1em,valign=m}{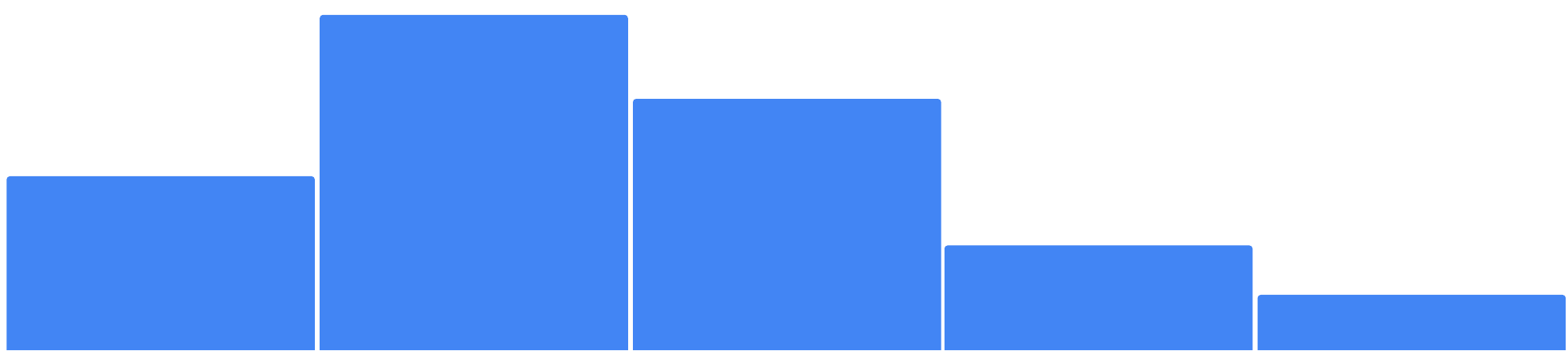} & 2.49 \\
    (\textbf{RQ2}) Dependabot often opens more PRs than I can handle.  & 
    \adjustimage{height=1.1em,valign=m}{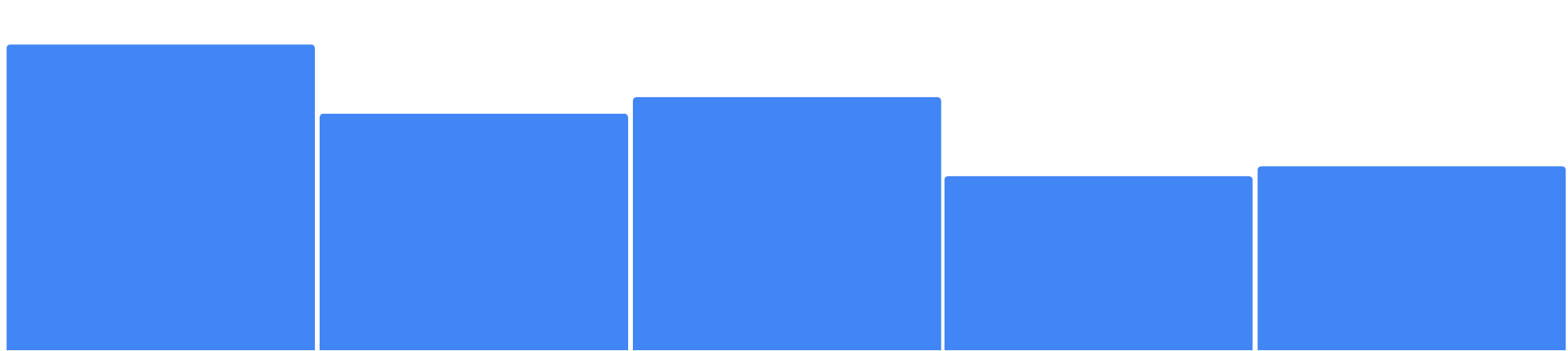} & 2.73 \\
    \midrule
    (\textbf{RQ3}) Compatibility scores are often available in Dependabot PRs.  & 
    \adjustimage{height=1.1em,valign=m}{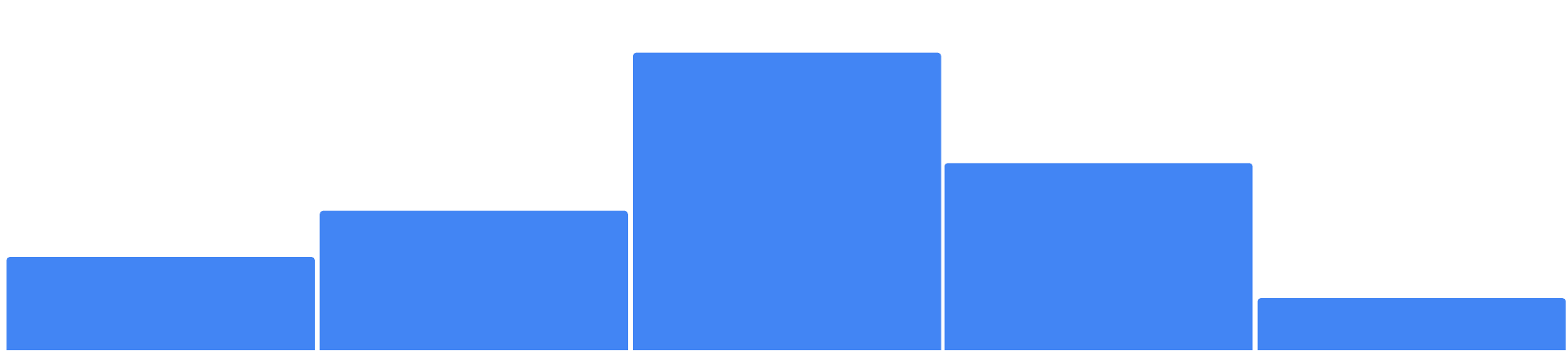}  & 2.95 \\
    (\textbf{RQ3}) If a compatibility score is available, it is effective in indicating whether the update will break my code.  & 
    \adjustimage{height=1.1em,valign=m}{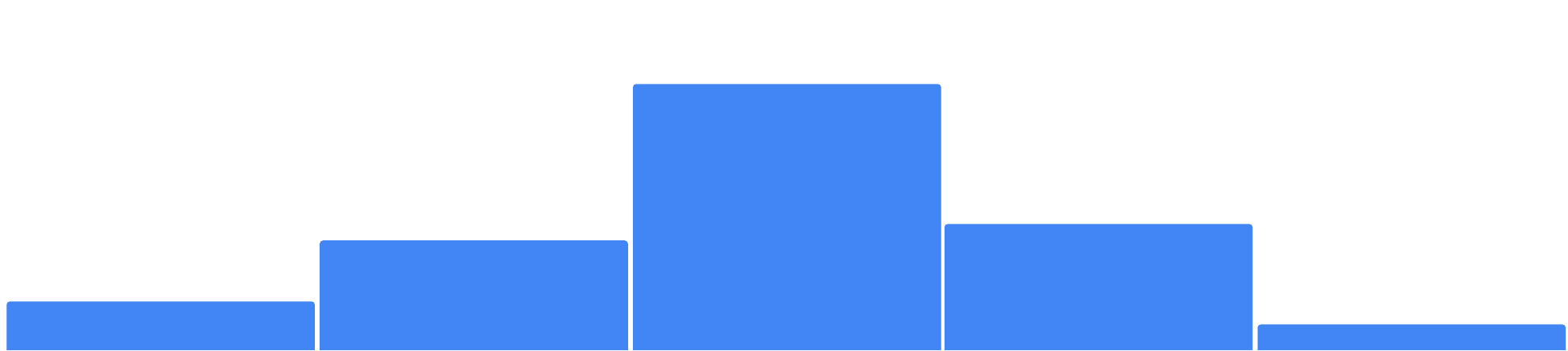} & 2.95 \\
    \midrule
    (\textbf{RQ4}) Dependabot can be configured to fit the needs of my project.  &
    \adjustimage{height=1.1em,valign=m}{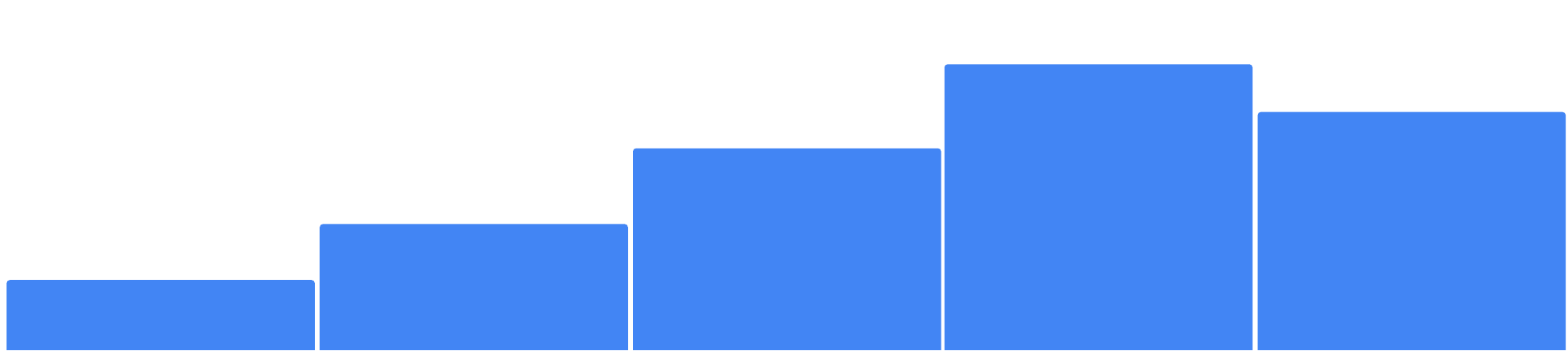} & 3.54 \\
    (\textbf{RQ4}) I configure Dependabot to make it less noisy (i.e., only update certain dependencies, scan less frequently, etc.)  & 
    \adjustimage{height=1.1em,valign=m}{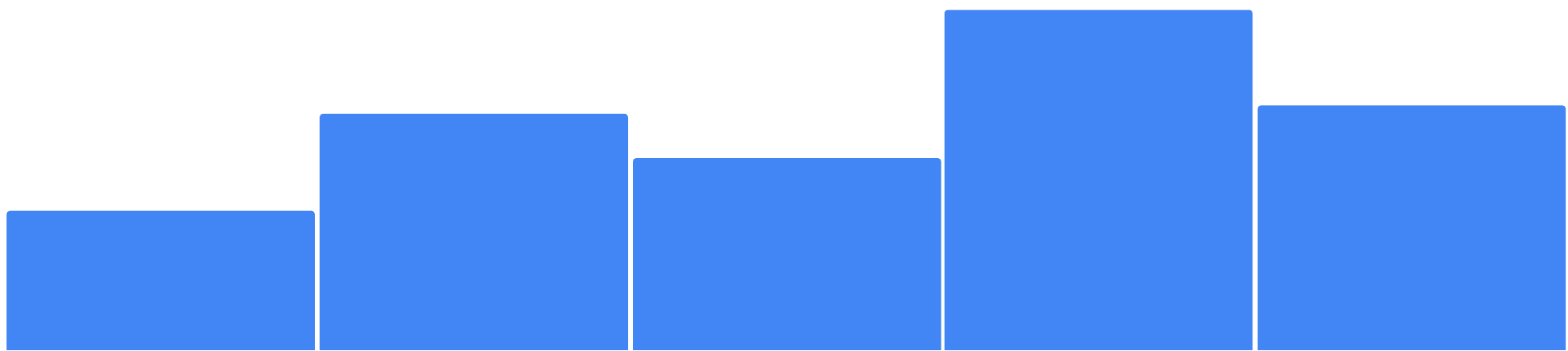} & 3.27 \\
    \midrule
    \multicolumn{3}{l}{\textbf{Multiple Choice Questions} } \\
    \midrule
    (\textbf{RQ5}) Are your GitHub repositories still using Dependabot for automating version updates? & \adjustimage{height=1.1em,valign=t}{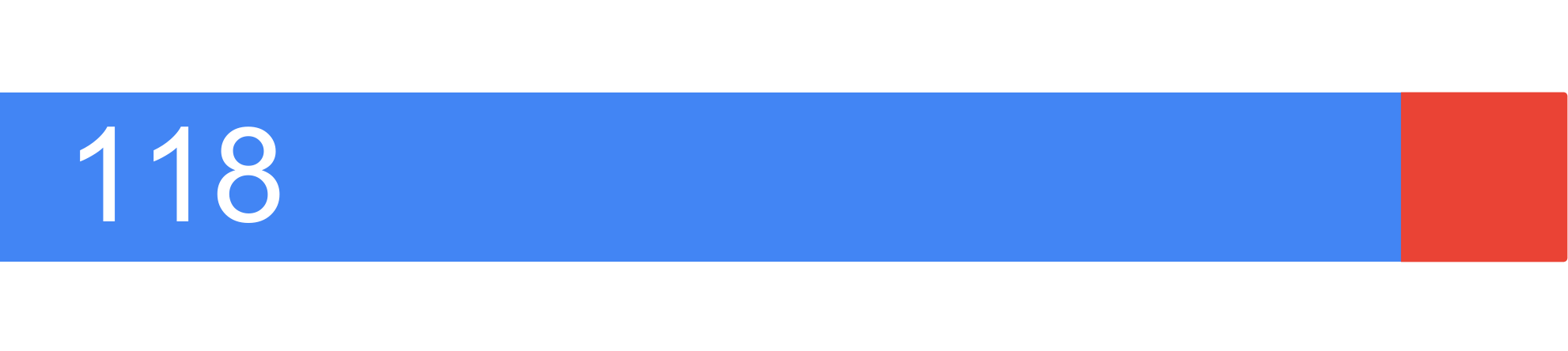} & 0.89 \\
    (\textbf{RQ5}) If not, why? & \multicolumn{2}{c}{(Results in \S~\ref{sec:rq5})} \\
    \midrule
    \multicolumn{3}{l}{\textbf{Open-Ended Questions}$^*$ } \\
    \midrule
    \makecell[l]{
        (\textbf{RQ5}) Regardless of current availability, what are the features you want most for a bot that updates dependencies?
        \\
        Do you have any further opinions or suggestions? 
    } & \multicolumn{2}{c}{ (Results in \S~\ref{sec:rq5}) }\\
  \bottomrule
  \end{tabularx}
  \begin{tablenotes}
    \item $^*$ Where appropriate, we also use evidence from open-ended question responses to support the results in \textbf{RQ1} - \textbf{RQ4}.
  \end{tablenotes}
  \end{threeparttable}
\vspace{-4mm}
\end{table*}

\textbf{PR Collection.}
We use GitHub REST API~\cite{GitHubAPI} and a web scraper to find all Dependabot PRs (before February 14, 2022) in the projects and collect PR statistics, CI test results, and timeline events.
By leveraging a distributed pool of Cloudflare workers~\cite{CloudflareWorker}, this web scraper empowers us to bypass the limitation of GitHub APIs (which is unhandy for collecting CI test results for PRs) and retrieve PR events and CI test results at scale.
The PR body can tell which dependency this PR is updating, its current version, and its updated version.
By the end of this stage, we obtain 540,665 Dependabot PRs (71.1\% with a CI test result), updating 15,390 dependencies between 167,841 version pairs.

Our next task is to identify security updates from all of the PRs created by Dependabot. 
However, Dependabot is no longer labeling security updates due to security reasons. 
Instead, Dependabot is showing a banner on the PR web page which is only visible to repository administrators by default \cite{DependabotAlertAccess}. 
Therefore, we choose to construct a mirror of the GitHub security advisory database~\cite{GitHubAdvisory} and identify security PRs ourselves by checking whether the PR updates a version with a vulnerability entry at the time of PR creation.
More specifically, we identify a PR to be a security update PR if: 1) the dependency and its current version matches a vulnerability in the GitHub security advisory database; 2) the updated version is newer than the version that fixes this vulnerability (i.e., no vulnerability after update); 3) the PR is created after the vulnerability disclosure in CVE.
Eventually, we identify 37,313 security update PRs (6.9\%) from the 540,665 Dependabot PRs in total.

\textbf{Dataset Overview.}
As illustrated in Table~\ref{tab:projectstats}, projects in our dataset are mostly engineered, popular GitHub projects with a large code base, active maintenance, rich development history, and frequent Dependabot usage.
We notice a long-tail distribution in the metrics concerning the size of the project, i.e., number of contributors, lines of code, and commit frequency, which is expected and common in most mining software repository (MSR) datasets~\cite{goeminne2011evidence, DBLP:journals/tse/ZhangZMJ21, 2022ICPC-ReleaseNote}. 

Most (44.1\%) projects in our dataset utilize the npm package ecosystem, followed by Maven (12.3\%), PyPI (11.7\%), and Go modules (7.8\%).
Among the Dependabot PRs, those that update npm packages constitute even a higher portion (64.9\%), followed by PyPI (8.9\%), Go modules (4.3\%), Bundler (3.9\%), and Maven (3.9\%), as packages in the npm ecosystem generally evolve faster~\cite{DBLP:journals/ese/DecanMG19}.

Dependabot has opened hundreds of PRs for most of the projects (mean = 304, median = 204), even up to thousands for some of them.
This likely indicates a high workload for project maintainers.
In terms of the most updated dependencies, it is not surprising that all top five comes from npm: \Code{@types/node} (29,352 PRs), \Code{eslint} (13,193 PRs), \Code{@typescript-eslint/parser} (11,833 PRs),  \Code{@typescript-eslint/eslint-plugin} (10,917 PRs), and \Code{webpack} (9,484 PRs). 
However, all these packages are mainly used as \Code{devDependencies} (static typing, linters, and module bundlers), which are typically only used for development but not in production.
Since the necessity of updating \Code{devDependencies} remains controversial~\cite{DBLP:journals/ese/KulaGOII18, DependabotDepdev}, Dependabot's frequent and massive updates to \Code{devDependencies} may be the first alarming signal of causing noise and notification fatigue to developers.


\subsection{Developer Survey}

To triangulate the results from data analysis, we additionally design and conduct a survey with developers from the 1,823 projects.\footnote{The survey has been approved by the Ethics Committee of Key Laboratory of High Confidence Software Technology, Ministry of Education (Peking University) under Grant No. CS20220011.}
The survey is summarized in Table~\ref{tab:survey} and consists of 11 5-point Likert scale~\cite{likert1932technique} questions (for \textbf{RQ1} - \textbf{RQ4}), two multiple choice questions, and two open-ended questions for collecting developers' desired features for dependency management bot, plus their further opinions if any (for \textbf{RQ5}).
To locate survey candidates, we find developers that are authors of commits that deprecate Dependabot, the most active respondents to its PRs, project owners, or the most active contributor in each project.
For each developer, we retrieve their email addresses from \Code{git} commits and exclude invalid emails (e.g., emails containing \Code{noreply}).
We get 1,295 developers after manually merging their identities.
Among them, we successfully deliver 1,226 emails and get 131 responses within three weeks (response rate 10.7\%).
This response rate represents a good maximum sample size~\cite{SampleSize} and is comparable to previous SE surveys~\cite{DBLP:conf/sigsoft/TanZS20, DBLP:journals/tse/ZhangZMJ21}.

We have carefully taken ethical considerations into account when designing and executing our survey. We only select a few survey candidates in each project (0.67 on average, one at most) who are most likely to be familiar with Dependabot and/or making important managerial decisions (i.e., adopting and deprecating tools like Dependabot).
For each candidate, we send personalized emails to them (with information about how they used Dependabot), to avoid being perceived as spam.
We try our best to follow common survey ethics~\cite{SurveyEthics}, e.g., clearly introducing the purpose of this survey, being transparent about what we will do to their responses, etc.
To increase the chance of getting a response and to contribute back to the open-source community, we offer to donate \$5 to an open-source project of the respondents' choice if they opt in.
Therefore, we believe we have done minimal harm to the open-source developers we have contacted, and the results we get about Dependabot far outweigh the harm. 
In fact, we get several highly welcoming responses from the survey participants, such as: 1) \textit{keep up the good work!} 2) \textit{If you would like to consult more, just ping me on <email>...Cheers!}

The bottom half of Table~\ref{tab:projectstats} summarizes the demographics of the 131 survey respondents, showing that they are highly experienced with both Dependabot (a median of 410 interactions) and open source development (five to 15 years of experience, hundreds of commits, and many followers).

\vspace{-0.5mm}

\section{Methods and Results}
\label{sec:results}

\subsection{RQ1: Technical Lag}
\vspace{-0.5mm}

\subsubsection{Repository Analysis Methods}
\label{sec:rq1-method}

We evaluate the effectiveness of Dependabot version updates by comparing the project technical lag at two time points: the day of Dependabot adoption ($T_0$) and 90 days after adoption (i.e., $T_0+90$). 
We choose 90 days as the interval to avoid the influence of deprecations 
as more than 85\% of them happen 90 days after adoption.
Since technical lag naturally increases over time~\cite{DBLP:journals/smr/ZeroualiMGDCR19, DBLP:conf/icsm/DecanMC18},
we include an additional time point for comparison: 90 days before adoption (i.e., $T_0-90$).

For a project $p$ at time $t \in \{T_0 - 90, T_0, T_0 + 90\}$, we denote all its direct dependencies as $\textbf{deps}(p, t)$ and define the technical lag of project $p$ at time $t$ as:
$$
\mathbf{techlag}(p, t)=\frac{\sum_{d\in \mathbf{deps}(p, t)}\mean\left(0, t_{\text{latest}}(d) - t_{\text{adopted}}(d)\right)}{|\mathbf{deps}(p, t)|}
$$
Here $t_{\text{latest}}(d)$ denotes the release time of $d$'s latest version at time $t$ and $t_{\text{adopted}}(d)$ denotes the release time of $d$'s adopted version. 
We use $\max$ to guard against the occasional case of $t_{\text{latest}}(d) < t_{\text{adopted}}(d)$ (e.g., developers may continue to release \Code{0.9.x} versions after the release of \Code{1.0.0}). 

This technical lag definition is inspired by Zerouali et al.~\cite{DBLP:journals/smr/ZeroualiMGDCR19} but with several adjustments. 
First, we use only their time-based variant instead of their version-based variant because cross-project comparisons would not be intuitive using the latter.
Second, we use the mean value of all dependencies instead of maximum or median as the overall technical lag, because we intend to measure the overall effectiveness of Dependabot for both keeping most dependencies up-to-date and eliminating the most outdated ones.

We exclude projects with an age of fewer than 90 days at Dependabot adoption and projects that deprecate Dependabot within 90 days.
We also exclude projects that migrate from Dependabot Preview since they may introduce bias into results.
Since the computation of technical lag based on dependency specification files and version numbers requires non-trivial implementation work for each package ecosystem, we limit our analysis on JavaScript/npm, the most popular ecosystem in our dataset. 
We further exclude projects with no eligible npm dependencies configured for Dependabot in $T_0-90$, $T_0$, or $T_0 + 90$.
After all the filtering, we retain 613 projects for answering RQ1.

We adopt the Regression Discontinuity Design (RDD) framework to estimate the impact of adopting Dependabot on project technical lags.
RDD uses the level of discontinuity before/after an intervention to measure its effect size while taking the influence of an overall background trend into consideration. 
Given that technical lag tends to be naturally increasing over time~\cite{DBLP:conf/icsm/DecanMC18, DBLP:conf/icsr/ZeroualiCMRG18, DBLP:journals/smr/ZeroualiMGDCR19}, RDD is a more appropriate statistic modeling approach for our case compared with hypothesis testing approaches (e.g., one-side Wilcoxon rank-sum tests). 
Following previous SE works that utilized RDD~\cite{DBLP:conf/kbse/ZhaoSZFV17, DBLP:conf/wcre/CasseeVS20}, 
we use sharp RDD, i.e., segmented regression analysis of interrupted time series data.
We treat project-level technical lag as a time series function, 
compute the technical lag for each project every 15 days from $T_0-90$ to $T_0+90$,
use ordinary least square regression to fit the RDD model,
and watch for the presence of discontinuity at Dependabot adoption, formalized as the following model:
\begin{multline*}
    y_i = \alpha + \beta \cdot time_i + \gamma \cdot intervention\\
     + \theta \cdot time\_after\_intervention_i + \sigma _i
\end{multline*}

Here $y_i$ denotes the output variable (i.e., technical lag for each project in our case);
$time$ stands for the number of days from $T_0-90$;
$intervention$ binarizes the presence of Dependabot (0 before adopting Dependabot, 1 after adoption);
$time\_after\_intervention$ counts the number of days from $T_0$ (0 when $T_0-90 \le time < T_0$).



\subsubsection{Repository Analysis Results}

\begin{table}[t]
  \scriptsize
  \centering
  \caption{Technical Lag (days) for 613 npm Projects}
\vspace{-2mm}
  \label{tab:techlags}
    \begin{tabular}{lrrc}
    \toprule
    Metric & Mean & Median & Distribution \\
    \midrule 
    $\mathbf{techlag}(p, T_0-90)$ & 73.68 & 16.27 & \adjustimage{height=0.47cm,valign=m}{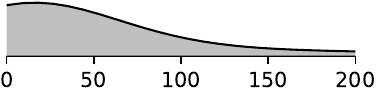}\\
    \MyIndent$\Delta$ in Between & -24.96 & 0.00 &  \adjustimage{height=0.47cm,valign=m}{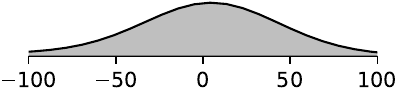}\\
    $\mathbf{techlag}(p, T_0)$ & 48.99 & 13.96 &  \adjustimage{height=0.47cm,valign=m}{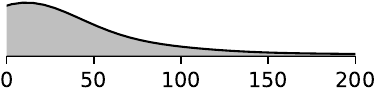}\\
    \MyIndent$\Delta$ in Between & -23.61 & -0.61 & \adjustimage{height=0.47cm,valign=m}{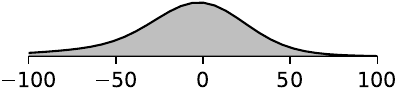}\\
    $\mathbf{techlag}(p, T_0+90)$ & 25.38 & 3.62 & \adjustimage{height=0.47cm,valign=m}{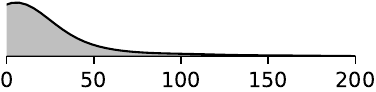}\\
    \bottomrule
    \end{tabular}
\end{table}

\begin{figure}[t]
  \footnotesize
  \centering
  \includegraphics[width=\linewidth]{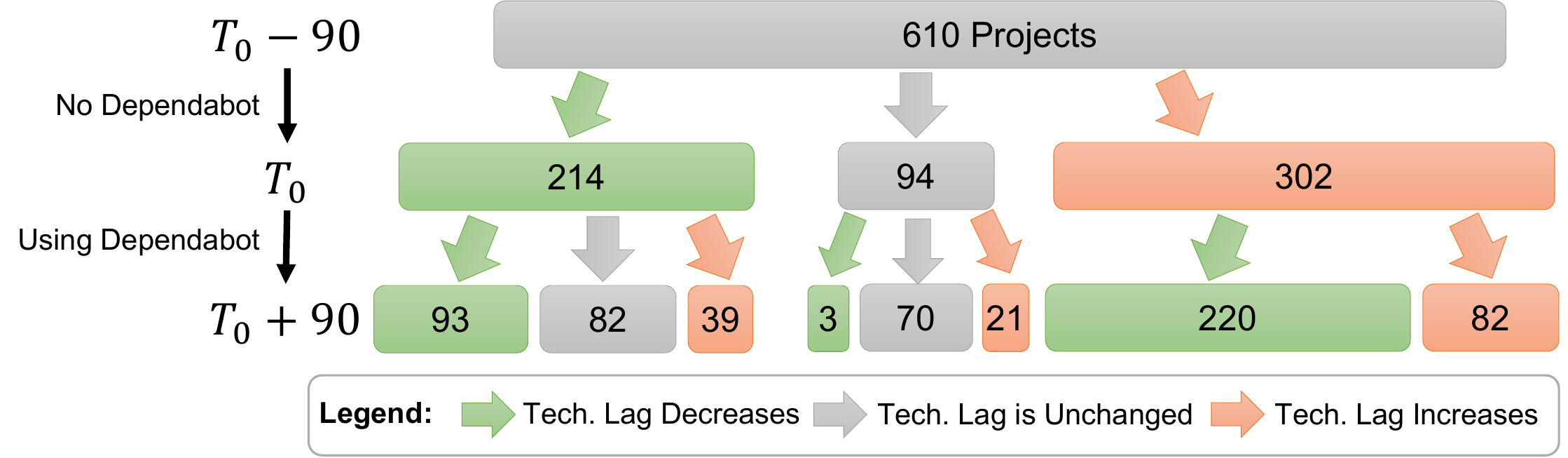}
\vspace{-3mm}
  \caption{
  Project-level technical lag changes ($T_0 - 90$,$T_0$,$T_0+90$).
  }
  \label{fig:lagchange}
\vspace{-3mm}
\end{figure}

We present technical lags and their delta between time points in Table~\ref{tab:techlags}.
We plot diagrams in Figure~\ref{fig:lagchange} to reflect how different projects increase/decrease their technical lag from $T_0-90$ to $T_0+90$.
The first surprising fact we notice is that the technical lag of approximately one-third (216/613) of projects is already decreasing between $T_0-90$ and $T_0$, even if technical lag tends to increase over time~\cite{DBLP:journals/smr/ZeroualiMGDCR19, DBLP:conf/icsm/DecanMC18}. 
This indicates that these projects are already taking a proactive dependency update strategy even before adopting Dependabot.
On the other hand, for about half (303/613) of the projects, the technical lag increases prior to Dependabot adoption, and 94 projects keep the technical lag unchanged.
For all projects, the mean and median technical lag at $T_0-90$ is 73.68 and 16.27 days, respectively; they decrease at $T_0$ to 48.99 and 13.96 days, respectively; at $T_0$, 159 (25.9\%) of the 613 projects have already achieved a zero technical lag.

Between $T_0$ and $T_0 + 90$, projects lower their technical lag even further from a mean of 48.99 days and a median of 13.96 days to a mean of 25.38 days and a median of 3.62 days.
Among the 303 projects with an increasing technical lag between $T_0-90$ and $T_0$, about two-thirds (220) of them see a decrease after adopting Dependabot; among the 216 projects with decreasing technical lag, nearly half (94) of them see a decrease. 
More than one-third (219, 35.7\%) of projects achieve completely zero technical lag 90 days after Dependabot adoption.
Although there are still some increases, the magnitude is much smaller (e.g., 75\% quantile of only $+1.75$ days between $T_0$ and $T_0+90$ compared with 75\% quantile of $+14.37$ days between $T_0-90$ and $T_0$).


\begin{table}[b]
  \scriptsize
  \centering
  \vspace{-3mm}
  \caption{
  The Estimated Coefficients and Significance Levels for the RDD Model We Fit (Section~\ref{sec:rq1-method}).
  }
  \vspace{-2mm}
  \begin{threeparttable}
  \begin{tabular}{lrrrr}
  \toprule
    Feature & Coef. & Std. Err. & $t$ & $p$ \\
    \midrule
    Intercept* & 66.5209 & 4.595 & 14.477 & 0.000 \\
    $intervention$* & -31.2137 & 5.694 & -5.306 & 0.000 \\
    $time$ & -0.0743 & 0.079 & -0.945 & 0.345 \\
    $time\_after\_intervention$ & -0.1011 & 0.100 & -1.008 & 0.314 \\
  \bottomrule
  \end{tabular}
  \begin{tablenotes}
      \item[$*$] $p < 0.001$
  \end{tablenotes}
  \end{threeparttable}
  \label{tab:rdd}
\end{table}

\begin{figure}[t]
  \centering
  \includegraphics[width=0.8\linewidth]{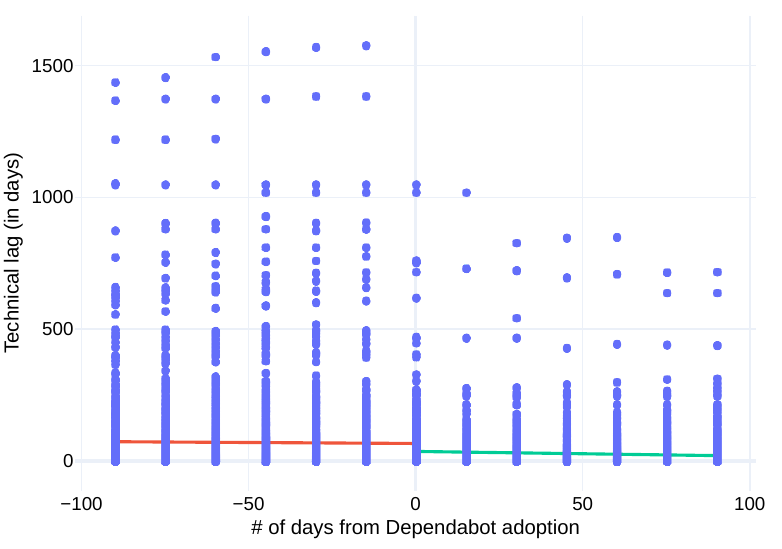}
  \vspace{-2mm}
  \caption{
  Project-level technical lag changes (from $T_0 - 90$ to $T_0+90$).
  Red and green lines represent the liner-fitted technical lag before/after Dependabot adoption respectively.
  }
  \label{fig:rdd}
\end{figure}

Table~\ref{tab:rdd} shows that the regression variable $intervention$ has a statistically significant negative coefficient ($coef. = -31.2137$, $p < 0.001$),
indicating the adoption of Dependabot might have reduced technical lag and kept dependencies up-to-date in the sampled 613 projects.
A more straightforward look at this trend can be observed in Figure~\ref{fig:rdd}:
at $T_0$, project-level technical lag has a noticeable decrease,
and there is a discontinuity between the liner-fitted technical lag before/after adoption.
$time$ and $time\_after\_intervention$ have negative coefficients, echoing with our earlier findings: 
the technical lag of sampled projects is already on decrease before Dependabot adoption 
and the introduction of Dependabot adds up to this decreasing trend.
However, both of the coefficients are not comparable to that of $intervention$ and are not statistically significant ($p > 0.3$).

\subsubsection{Triangulation from Survey}
\label{sec:rq1-validation}

Most developers agree that Dependabot is helpful in keeping their project dependencies up-to-date: 55.8\% responded with \textbf{Strongly Agree} and 35.7\% with \textbf{Agree} (Table~\ref{tab:survey}). 
As noted by one developer: \textit{Dependabot does a great job of keeping my repositories current}.
This is because Dependabot serves well as an automated notification mechanism that tells them about the presence of new versions and pushes them to update their dependencies.
As mentioned by two developers: 1) \textit{Dependabot is a wonderful way for me to learn about major/minor updates to libraries.}
2) \textit{Dependabot can be a bit noisy, but it makes me aware of my dependencies.}

However, some of the developers do not favor using Dependabot for automating dependency updates but only use Dependabot as a way of notification.
For example:
1) \textit{We just use it for notifications about updates, but do them manually and check if anything broke in the process.}
2) \textit{I am just using Dependabot to tell me if there is something to update and then update all in a single shot with plain package managers.}

This indicates that they do not trust the reliability of Dependabot for automating updates and they do not think the current design of Dependabot can help them reduce the manual workload of updates.
As an example, one developer states that: \textit{Dependency management is currently much easier just utilizing yarn/npm. We use Dependabot merely because it has been recommended, but updating dependencies was faster when I solely used the command line.}


One developer suggests that using Dependabot only for update notifications has become such a common use case that they would prefer a dedicated, less noisy tool solely designed for this purpose:
\textit{It (Dependabot) becomes more like an update notification, i.e. I'm leveraging only half of its capability. Could there be something designed solely for this purpose? Less invasive, more informative, and instead of creating a PR for every package's update, I would like to see a panel-style hub to collect all the information for me to get a better overview in one place.}

\begin{small}
\begin{result-rq}{Findings for RQ1:}
90 days after adopting Dependabot, projects decrease their technical lag from an average of 48.99 days to an average of 25.38 days.
35.7\% of projects achieve zero technical lag 90 days after adoption.
The adoption of Dependabot is a statistically significant intervention as indicated by RDD. 
Developers agree on its effectiveness in \textit{notifying} updates, but question its effectiveness in \textit{automating} updates.
\end{result-rq}
\end{small}
\vspace{-1mm}

\subsection{RQ2: Developers' Response to Pull Requests}


\subsubsection{Repository Analysis Methods}
Inspired by prior works~\cite{DBLP:conf/msr/AlfadelCSM21, DBLP:conf/botse-ws/WyrichGHM21}, we use the following metrics to measure the receptiveness (i.e., how active developers merge) and responsiveness (i.e., how active developers respond) of Dependabot PRs:
\begin{itemize}[leftmargin=15pt]
    \item \textbf{Merge Rate}: The proportion of merged PRs.
    \item \textbf{Merge Lag}: The time it takes for a PR to be merged. 
    \item \textbf{Close Lag}: The time it takes for a PR to be closed (i.e., not merged into the project code base).
    \item \textbf{Response Lag}: The time it takes for a PR to have human interactions, including any observable action in the PR's timeline, e.g., adding a label or assigning a reviewer.
\end{itemize}
The merge rate is intended to measure receptiveness and the latter three are intended to measure responsiveness.

We assume that results may differ for PRs in different groups.
We expect that 
1) developers are both more receptive and more responsive to security updates due to their higher priority of eliminating security vulnerabilities; 
and 2) projects that use Dependabot version update (i.e., contain \Code{dependabot.yml}s) are more responsive to Dependabot PRs. 
To verify our expectations, we divide PRs into three groups:
\begin{itemize}[leftmargin=15pt]
    \item \textbf{regular}: Dependabot PRs that update a package to its latest version when the old version does not contain any known security vulnerabilities.
    \item \textbf{sec/conf}: Security PRs that update a package with vulnerabilities to its patched version and are opened when the project has a \Code{dependabot.yml} file in its repository (i.e., using Dependabot version update). 
    \item \textbf{sec/nconf}: Security PRs opened when the project does not have a \Code{dependabot.yml} file in its repository. These PRs are opened either \textit{before} the adoption or \textit{after} the deprecation of Dependabot version update.
\end{itemize}
We examine the significance of inter-group metric differences with unpaired Mann-Whitney tests and Cliff's delta ($\delta$).
Following Romano et al.~\cite{romano2006appropriate}, we consider the effect size as negligible for $|\delta| \in [0, 0.147)$, small for $|\delta| \in [0.147, 0.33)$, medium for $|\delta| \in [0.33, 0.474)$, and large otherwise.


\subsubsection{Repository Analysis Results}

\begin{table}[t]
  \footnotesize
  \centering
  \caption{
  PR Statistics in Different Groups.
  All lags are measured in days. 
  $\bar{x}$ represents the mean and $\mu$ represents the median over all PRs in this group.
  }
\vspace{-2mm}
  \label{tab:lags}
    \scriptsize
    \begin{tabular}{lrrr}
    \toprule
    Statistics & \textbf{regular} & \textbf{sec/conf} & \textbf{sec/nconf} \\
    \midrule 
    \# of PRs & 502,752 & 13,406 & 23,907\\
    Merge Rate & 70.13\% & 73.71\% & 76.01\% \\
    Merge Lag & $\bar{x}$=1.76, $\mu$=0.18 & $\bar{x}$=3.45, $\mu$=0.18 & $\bar{x}$=8.15, $\mu$=0.76\\
    Close Lag & $\bar{x}$=8.63, $\mu$=3.00 & $\bar{x}$=14.42, $\mu$=5.00 & $\bar{x}$=26.83, $\mu$=5.71\\
    Resp. Lag & $\bar{x}$=2.27, $\mu$=0.17 & $\bar{x}$=3.74, $\mu$=0.17 & $\bar{x}$=8.59, $\mu$=0.51\\
    \bottomrule
    \end{tabular}
\vspace{-3mm}
\end{table}

Table~\ref{tab:lags} shows the PR statistics we obtain for each group.
The high merge rates ($>$70\%) indicate the projects are highly receptive to Dependabot PRs regardless of whether they are security-related.
They are more receptive to security PRs: their merge rate is 74.53\%, even higher than 65.42\% reported on Dependabot Preview security updates~\cite{DBLP:conf/msr/AlfadelCSM21}.
This may be because projects welcome security updates even more, or just because the projects we selected are such.



Alfadel et al.~\cite{DBLP:conf/msr/AlfadelCSM21} find that Dependabot security PRs take longer to close than to merge. 
Our data illustrate a similar story: \textbf{regular} Dependabot PRs take a median of 0.18 days ($\approx$ four hours) to merge and a median of 3.00 days to close.
The difference is statistically significant with a large effect size ($p < 0.001$, $\delta = 0.91$).

The response lag, however, does not differ much from the merge lag in all groups, which confirms the timeliness of developers' response towards Dependabot PRs. 
We observe human activities in 360,126 (72.2\%) Dependabot PRs, among which 280,276 (77.8\%) take less than one day to respond.
However, this also indicates an inconsistency between fast responses and slow closes. 
As a glance at what caused this inconsistency, we sample ten closed PRs with developers' activities before closing and inspect their event history. 
We find 9 out of 10 PRs are closed by Dependabot itself, for the PR being obsolete due to the release of a newer version or a manual upgrade (similar to the observation by Alfadel et al.~\cite{DBLP:conf/msr/AlfadelCSM21}).
Activities are development-related (e.g., starting a discussion, assigning reviewers) in 5 PRs, while the rest are interactions with Dependabot (e.g., \Code{@dependabot rebase}).

Surprisingly, security PRs require a longer time to merge ($p < 0.001$, $\delta = 0.87$), close ($p < 0.001$, $\delta = 0.72$), and respond ($p < 0.001$, $\delta = 0.87$) with large effect sizes, regardless of whether the project is using Dependabot version update.
Though Dependabot version update users do process security updates quicker (at least merge lag and response lag are noticeably shorter), this difference is not significant with negligible or small effect sizes ($\delta \le 0.23$).


\subsubsection{Triangulation from Survey}
\label{sec:rq2-validation}


In general, developers agree that Dependabot PRs do not require much work to review and merge (34.1\% \textbf{Strongly Agree}, 40.3\% \textbf{Agree}, 14.0\% \textbf{Neutral}).

We find that they follow two different patterns of using Dependabot.
One pattern is to rapidly merge the PR if the tests pass and manually perform the update by hand otherwise (65.2\% \textbf{Strongly Agree}, 19.7\% \textbf{Agree}, 9.1\% \textbf{Neutral}).
In the latter case, they will respond to the Dependabot PR slower, or let Dependabot automatically close the PR after the manual update (36.4\% \textbf{Strongly Agree}, 26.5\% \textbf{Agree}, 20.5\% \textbf{Neutral}).
For example: \textit{I almost never have to look at Dependabot PRs because I have tests, and 99.99\% of PRs are merged automatically. Rarely (when dependency changes API for example) I have to manually add some fixes/updates...}
As mentioned in Section~\ref{sec:rq1-validation}, another pattern is to use Dependabot PRs solely as a way of notification and always perform manual updates.
Both cases contribute to the much larger close lag we observe in Dependabot PRs.

In terms of security updates, most developers do handle security PRs with a higher priority (56.7\% \textbf{Strongly Agree}, 16.3\% \textbf{Agree}, 14.0\% \textbf{Neutral}), but they do not think security PRs require more work to review and merge (19.4\% \textbf{Totally Disagree}, 36.4\% \textbf{Disagree}, 26.4\% \textbf{Neutral}). 
One possible explanation for the slower response, merge, and close of security PRs is that developers consider some security vulnerabilities as irrelevant to them: \textit{I want it (Dependabot) to ignore security vulnerabilities in development dependencies that don't actually get used in production}.

Developers have a mixed opinion on whether Dependabot opens more PRs than they can handle (15.9\% \textbf{Strongly Agree}, 15.2\% \textbf{Agree}, 22.0\% \textbf{Neutral}, 20.5\% \textbf{Disagree}, 26.5\% \textbf{Totally Disagree}).
Whether the PR workload introduced by Dependabot is acceptable may depend on other factors (e.g., the number of dependencies and how fast packages evolve), as indicated by two respondents: 
1) \textit{The performance of Dependabot or other similar bots could depend on the number of dependencies a project has. For smaller projects, with a handful of dependencies, Dependabot will be less noisy and usually safe as compared to large projects with a lot of dependencies.}
2) \textit{The utility of something like Dependabot depends heavily on the stack and number of dependencies you have. JS is much more noisy than Ruby, for example, because Ruby moves more slowly.}

\begin{small}
\begin{result-rq}{Findings for RQ2:}
>70\% of Dependabot PRs are merged with a median merge lag of four hours.
Compared with regular PRs, developers are less \textit{responsive} (more time to respond, close or merge) but more \textit{receptive} (higher merge rate) to security PRs.
Developers tend to rapidly merge PRs they consider ``safe'' and perform manual updates for the remaining PRs. 
\end{result-rq}
\end{small}

\subsection{RQ3: Compatibility Score}


\subsubsection{Repository Analysis Methods}

We explore the effectiveness of compatibility scores in two aspects: \textbf{Availability}, and \textbf{Correlation with Merge Rate}.

\textbf{1) Availability: }
We begin our analysis by understanding the data availability of compatibility scores, for they would not take effect if they are absent from most of the PRs.
For this purpose, we obtain compatibility scores from badges in PR bodies, which point to URLs defined \textit{per dependency version pair}.
That is, Dependabot computes one compatibility score for each dependency version pair $\langle d, v_1, v_2 \rangle$ and show the score to all PRs that update dependency $d$ from $v_1$ to $v_2$.
In case this computation fails, Dependabot generates an \Code{unknown} compatibility score for $\langle d, v_1, v_2 \rangle$.

Since compatibility scores are computed in a data-driven manner, we wonder if the popularity of the updated dependencies affects their availability.
As a quick evaluation, we sample 20 npm dependencies with more than one million downloads per week as representatives for popular dependencies. 
Next, we retrieve the release history of these dependencies by querying the npm registry API, retaining only releases that came available after January 1, 2020 (recall that all Dependabot PRs in our dataset are created after January 2020, Section~\ref{sec:datacollection}).
For the releases in each dependency, we get all possible dependency version pairs from a Cartesian product (1,629 in total) and query their compatibility scores from corresponding Dependabot URLs.

\textbf{2) Correlation with Merge Rate: }
In theory, if developers perceive compatibility scores as reliable, PRs with higher compatibility scores will be more likely to get merged.
To quantitatively evaluate this, we compare merge rates for PRs with different compatibility scores.
Since PRs that update the same version pair share the same score, we further utilize Spearman's $\rho$ to measure the correlation between a) compatibility score for a dependency version pair $\langle d, v_1, v_2 \rangle$, and b) merge rate for all PRs that update $d$ from $v_1$ to $v_2$.

As we will show in Section~\ref{sec:rq3-results}, compatibility scores are abnormally scarce.
Although we have reached Dependabot maintainers for explanations, they claim such information to be confidential and refuse to share any details.
We compute the number of CI test results for each dependency version pair and analyze their overall distribution to provide possible explanations for such scarcity.

\subsubsection{Repository Analysis Results}
\label{sec:rq3-results}

\begin{figure}[t]
\centering
\begin{subfigure}{0.5\linewidth}
\centering
    \includegraphics[width=0.95\linewidth]{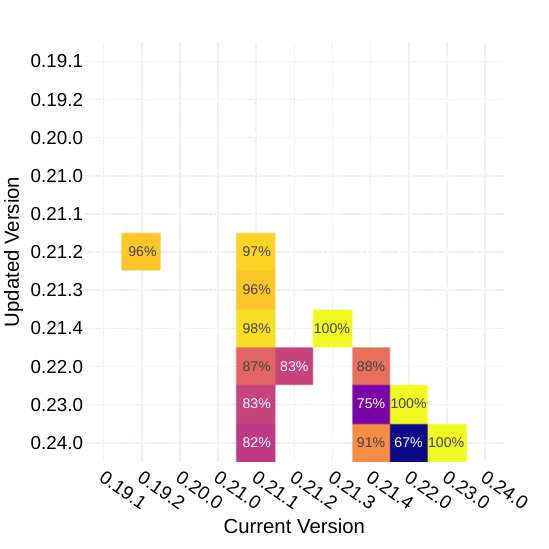}
    \caption{Compatibility Score}
    \label{fig:axios_compat}
\end{subfigure}%
\begin{subfigure}{0.5\linewidth}
\centering
    \includegraphics[width=0.95\linewidth]{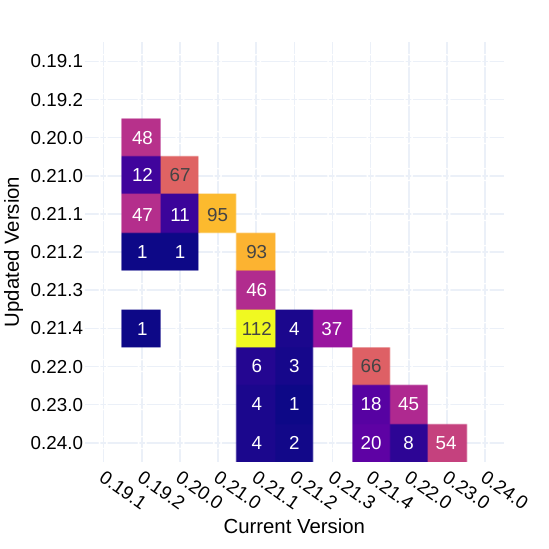}
    \caption{\# of CI Test Results}
    \label{fig:axios_numcicheck}
\end{subfigure}
\vspace{-1mm}
\caption{Distribution of compatibility scores and available CI test results over the version pairs of \Code{axios}.}
\label{fig:compatdist}
\end{figure}

\begin{table}[t]
  \footnotesize
  \centering
  \caption{Compatibility Score and PR Merge Rate}
\vspace{-2mm}
  \label{tab:compat}
    \begin{tabular}{lrr}
    \toprule
    Compatibility Score & \# of PRs & Merge Rate \\
    \midrule 
    \Code{unknown} & 485,501 & 69.96\% \\
    \Code{< 80\%} & 1,321 & 30.20\% \\
    \Code{< 90\%, >= 80\%} & 1,605 & 67.48\%\\
    \Code{< 95\%, >= 90\%} & 1,794 & 73.19\%\\
    \Code{< 100\%, >= 95\%} & 2,228 & 84.43\%\\
    \Code{== 100\%} & 10,303 & 80.30\%\\
    \bottomrule
    \end{tabular}
\vspace{-3mm}
\end{table}

\textbf{1) Availability: } 
Compatibility scores are extremely scarce: 
Only 3.4\% of the PRs and 0.5\% of the dependency version pairs have a compatibility score other than \Code{unknown}.
Merely 0.18\% of the dependency version pairs have a value other than \Code{100\%}.
Its scarcity does not become better even among the most popular npm dependencies:
1,604 (98.5\%) of the 1,629 dependency version pairs we sample only have a compatibility score of \Code{unknown}, 10 (0.6\%) have a compatibility score of \Code{100\%}, and 15 (0.9\%) have a compatibility score less than \Code{100\%}.
As an example, we plot a compatibility score matrix for \Code{axios}, which has the most (15) version pairs with compatibility scores, in Figure~\ref{fig:axios_compat}.

\textbf{2) Correlation with Merge Rate: }
We summarize the merge rates for PRs with different compatibility scores in Table~\ref{tab:compat}.
We can observe that for PRs with a compatibility score, a high score indeed increases their chance of being merged: if the score is higher than \Code{90\%}, developers are more likely to merge the PR.
By contrast, if the score is lower than \Code{80\%}, developers become very unlikely (30.20\%) to merge.
The Spearman’s $\rho$ between compatibility score and merge rate is $0.37$ ($p<0.001$), indicating a weak correlation according to Prion and Haerling's interpretation~\cite{spearmanRhoPrion}.

\begin{figure}[t]
  \centering
  \includegraphics[width=\linewidth]{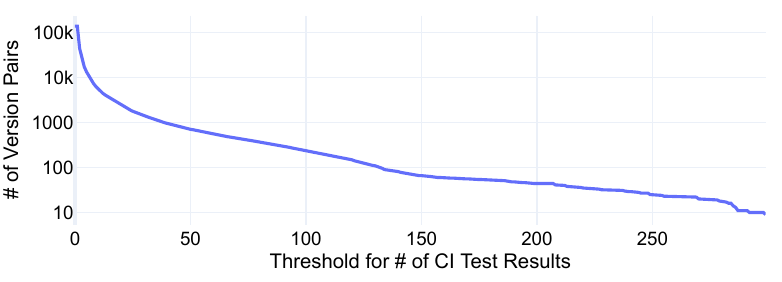}
\vspace{-5mm}
  \caption{
    \# of dependency version pairs with more than $x$ CI test results. Note that the $y$ axis is log-scale.
  }
\vspace{-3mm}
  \label{fig:checkspearman}
\end{figure}

Figure~\ref{fig:checkspearman} shows the number of dependency version pairs with more than $x$ CI test results.
We can observe an extreme Pareto-like distribution: for the 167,053 dependency version pairs in our dataset, less than 1,000 have more than 50 CI test results and less than 100 have more than 150 CI test results.
For the case of \Code{axios} (Figure~\ref{fig:axios_numcicheck}), the compatibility scores are indeed only available for version pairs with available CI test results.
It is hard to explain why the scores are missing even for some version pairs with many CI test results (e.g., the update from \Code{0.19.2} to \Code{0.20.0}), as we do not know the underlying implementation details.

\subsubsection{Triangulation from Survey}
\label{sec:rq3-validation}


Developers have diverging opinions on whether compatibility scores are available (7\% \textbf{Strongly Agree}, 24.8\% \textbf{Agree}, 38.8\% \textbf{Neutral}, 17.8\% \textbf{Disagree}, 11.6\% \textbf{Totally Disagree}) and whether compatibility scores are effective if they are available (4.7\% \textbf{Strongly Agree}, 21.7\% \textbf{Agree}, 45.7\% \textbf{Neutral}, 19.4\% \textbf{Disagree}, and 8.5\% \textbf{Totally Disagree}).
The answer distributions and the high number of \textbf{Neutral} responses likely indicate that many developers do not know how to rate the two statements~\cite{sturgis2014middle}, because compatibility scores are too scarce and most developers have not been exposed to this feature.
As replied by one developer: 
\textit{Compatibility scores and vulnerable dependencies detection are great, I use Dependabot a lot but was not aware they exist...(They) should be more visible to the user.} 
Another developer does express concerns that compatibility scores are not effective, saying that \textit{Dependabot's compatibility score has never worked for me.}

Further, several developers (6 responses in our survey) hold the belief that Dependabot only works well in projects with a high-quality test suite. For example:
\begin{enumerate}[leftmargin=15pt]
    \item \textit{Dependabot works best with a high test coverage and if it fails people it's likely because they have too little test coverage.}
    \item \textit{Dependabot without a good test suite is indeed likely too noisy, but with good tests and an understanding of the code base it is trivial to know whether an update is safe to update or not.}
\end{enumerate}

\begin{small}
\begin{result-rq}{Findings for RQ3:}
Compatibility scores are too scarce to be effective: only 3.4\% of PRs have a known compatibility score.
For those PRs with one, the scores have a weak correlation ($\rho = 0.37$) with the PR merge rate.
Its scarcity may be because most dependency version pairs do not have sufficient CI test results (i.e., a Pareto-like distribution) for inferring update compatibility.
As a result, developers think Dependabot only works well in projects with high-quality test suites.
\end{result-rq}
\end{small}
\vspace{-0.5mm}

\subsection{RQ4: Configuration}


\subsubsection{Repository Analysis Methods}

Dependabot offers tons of configuration options for integration with project workflows, such as who to review, how to write commit messages, how to label, etc.
In this research question, we only focus on the options related to \textit{notifications} because we expect them to be possible countermeasures against noise and notification fatigue.
More specifically, we investigate the following options provided by Dependabot:
\begin{enumerate}[leftmargin=15pt]
    \item \Code{schedule.interval}: This option is mandatory and specifies how often Dependabot scans project dependencies, checks for new versions, and opens update PRs. Possible values include \Code{"daily"}, \Code{"weekly"}, and \Code{"monthly"}.
    \item \Code{open-pull-requests-limit}: It specifies the maximum number of simultaneously open Dependabot PRs allowed in a project. The default value is five.
    \item \Code{allow}: It tells Dependabot to only update a subset of dependencies. By default, all dependencies are updated.
    \item \Code{ignore}: It tells Dependabot to ignore a subset of dependencies. By default, no dependency is ignored.
\end{enumerate}
The latter two options are very flexible and may contain constraints exclusive to some package ecosystems, e.g., allowing updates in production manifests or ignoring patch updates according to the semantic versioning convention~\cite{SemVer}. 

To understand developers' current practice of configuring Dependabot, we parse 3,921 Dependabot configurations from 1,588 projects with a \Code{dependabot.yml} in their current working tree.\footnote{
Note that 235 of the 1,823 projects do not have \texttt{dependabot.yml} in their current working tree which we will investigate in \textbf{RQ5}.
One project may depend on more than one package ecosystem (e.g., both npm and PyPI) and have separate configurations for each of them.
}
For \Code{schedule.interval} and \Code{open-pull-requests-limit}, we count the frequency of each value.
For \Code{allow} and \Code{ignore}, we parse different options and group them into three distinctive strategies:
\begin{enumerate}[leftmargin=15pt]
    \item \textbf{default}: allowing Dependabot to update all dependencies, which is its default behavior;
    \item \textbf{ignorelist}: configuring Dependabot to ignore a subset of dependencies;
    \item \textbf{allowlist}: configuring Dependabot to only update on a subset of dependencies.
\end{enumerate}

We further explore the modification history of Dependabot configurations to observe how developers use configuration as a countermeasure against noise in the wild.
For this purpose, we find all commits in the 1,823 projects that modified \Code{dependabot.yml} and extract eight types of configuration changes from file diffs:
\begin{enumerate}[leftmargin=15pt]
    \item \Code{+interval}: Developers increase \Code{schedule.interval}.
    \item \Code{-interval}: Developers decrease \Code{schedule.interval}.
    \item \Code{+limit}: Developers increase \scalebox{.95}[1.0]{\Code{open-pull-requests-limit}}.
    \item \Code{-limit}: Developers decrease \scalebox{.94}[1.0]{\Code{open-pull-requests-limit}}.
    \item \Code{+allow}: Developers allow some more dependencies to be automatically updated by Dependabot.
    \item \Code{-allow}: Developers no longer allow some dependencies to be automatically updated by Dependabot.
    \item \Code{+ignore}: Developers configure Dependabot to ignore some dependencies for automated update.
    \item \Code{-ignore}: Developers configure Dependabot to no longer ignore some dependencies for automated update.
\end{enumerate}

Finally, we analyze configuration modifications by time since Dependabot adoption. We mainly focus on the bursts of modification patterns, because bursts illustrate the lag from the developers' perception of noise to their countermeasures to mitigate the noise.

\subsubsection{Repository Analysis Results}
\label{sec:config-results}

\begin{figure}[t]
  \centering
  \includegraphics[width=\linewidth]{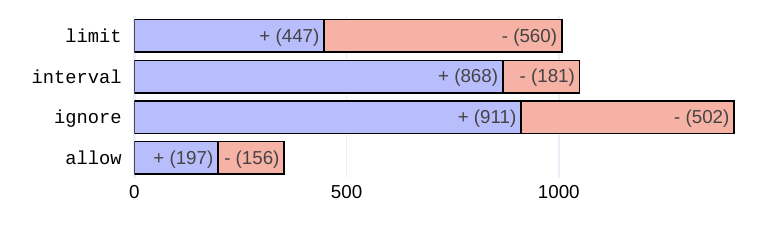}
\vspace{-6mm}
  \caption{Distribution of configuration modifications by type.}
  \label{fig:modtype}
\end{figure}

\begin{figure}[t]
\centering
\begin{subfigure}{0.5\linewidth}
\centering
    \includegraphics[width=0.9\linewidth]{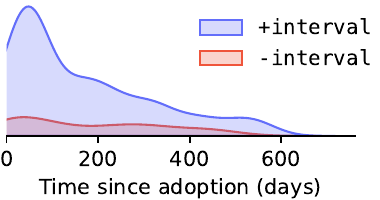}
    \vspace{-1mm}
    \caption{\Code{schedule.interval}}
    \vspace{4mm}
    \label{fig:modinterval}
\end{subfigure}%
\begin{subfigure}{0.5\linewidth}
\centering
    \includegraphics[width=0.9\linewidth]{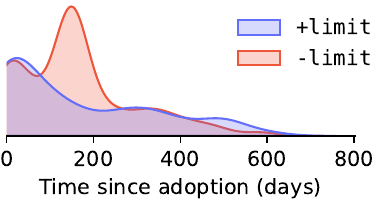}
    \vspace{-1mm}
    \caption{\scalebox{.98}[1.0]{\Code{open-pull-requests-limit}}}
    \vspace{4mm}
    \label{fig:modprlimit}
\end{subfigure}
\begin{subfigure}{0.5\linewidth}
\centering
    \includegraphics[width=0.9\linewidth]{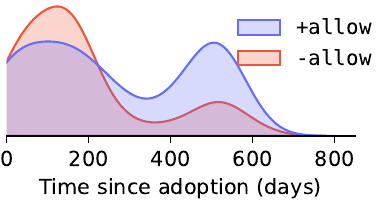}
    \vspace{-1mm}
    \caption{\Code{allow}}
    \label{fig:modallow}
\end{subfigure}%
\begin{subfigure}{0.5\linewidth}
\centering
    \includegraphics[width=0.9\linewidth]{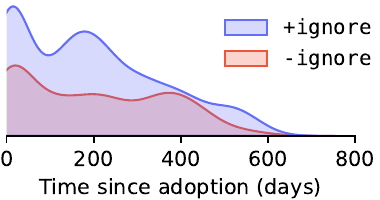}
    \vspace{-1mm}
    \caption{\Code{ignore}}
    \label{fig:modignore}
\end{subfigure}
\caption{Config. modifications since Dependabot adoption.}
\label{fig:confmod}
\vspace{-3mm}
\end{figure}

The current configurations of Dependabot show that most projects configure Dependabot toward a proactive update strategy: 
2,203 (56.2\%) of \Code{schedule.interval} are \Code{"daily"} while merely 276 (7.04\%) of them are a conservative \Code{"monthly"}.
1,404 (35.8\%) of the \Code{open-pull-requests-limit} configurations are higher above the default value while only a negligible proportion (2.3\%) is lower.
For \Code{allow} and \Code{ignore} options, most of the configurations (3,396, 86.7\%) adopt the \textbf{default} strategy, less (380, 9.7\%) use \textbf{ignorelist}, and a small proportion (50, 1.3\%) use \textbf{allowlist}.

The modifications tell us another story.
776 (42.57\%) of the 1,823 projects in our dataset have modified the Dependabot configuration options we study (e.g., update interval) and they contain 2.18 modification commits on average (median = 1.00).
Figure~\ref{fig:modtype} illustrates the proportion of each modification type, which shows that projects increase \Code{schedule.interval} and lower \Code{open-pull-requests-limit} more frequently than doing the opposite.
As demonstrated in Figure~\ref{fig:confmod}, projects can increase \Code{schedule.interval} any time after Dependabot adoption but more likely to reduce \Code{open-pull-requests-limit} only after several months of Dependabot usage.
\Code{schedule.interval} determines how often Dependabot bothers developers to a large extent, and we are seeing developers of 336 projects increasing it in 868 configurations. 
We further confirm this behavior as a countermeasure against noise from a real-life example where developers \textit{reduce the frequency to monthly to reduce noise}~\cite{dropbox/stone/pull/259}.
\Code{open-pull-requests-limit} quantifies the developers' workload on each interaction, which is also noise-related as indicated by a developers' complaint: \textit{Dependabot PRs quickly get out of hand}~\cite{tuist/tuist/pull/3155}.
If we focus on modifications that happen 90 days after Dependabot adoption, we find nearly two-thirds (62.5\%) of \Code{open-pull-requests-limit} changes belong to \Code{-limit}.
Our observations indicate the following phenomenon.
At the beginning of adoption, developers configure Dependabot to interact frequently and update proactively.
However, they later get overwhelmed and suffer from notification fatigue, which causes them to reduce interaction with Dependabot or even deprecate Dependabot (\textbf{RQ5}).
As an extreme case, one developer forces Dependabot to open only 1 PR at a time to reduce noise~\cite{ros-tooling/action-ros-ci/pull/663}.

Ignoring certain dependencies seems to be another noise countermeasure, for developers tend to add an ignored dependency more often than remove one (Figure~\ref{fig:modtype}).
For example, a commit says \textit{update ignored packages...so they are never automatically updated to stop noise}~\cite{justeat/httpclient-interception/commit/b337b5f}. 
However, we also observe cases where developers add ignored dependencies due to other intentions, such as handling breaking changes~\cite{asynkron/protoactor-dotnet/pull/1260} and preserving backward compatibility~\cite{Azure/bicep/commit/a06b0e4}.
For \Code{+allow} and \Code{-allow}, we observe an interesting burst of \Code{-allow} (Figure~\ref{fig:modallow}) earlier but more \Code{+allow} dependencies later, but we do not find any evidence explaining such trend.


\subsubsection{Triangulation from Survey}

Although more than half of respondents think Dependabot can be configured to fit their needs (25.6\% \textbf{Strongly Agree} and 30.2\% \textbf{Agree}), some do not (7.8\% \textbf{Totally Disagree} and 14\% \textbf{Disagree}).
As a peek into this controversy, one developer says, \textit{I think people that complain about how noisy it is (I've seen a lot of this) just don't configure things correctly.}

More than half (50.4\%) of respondents have configured Dependabot to make it less noisy, but roughly one-third (32.6\%) have not (21.2\% \textbf{Strongly Agree}, 29.5\% \textbf{Agree}, 16.7\% \textbf{Neutral}, 20.5\% \textbf{Disagree}, 12.1\% \textbf{Totally Disagree}).
It is possible that the default configurations of Dependabot only work for projects with a limited number of dependencies and these dependencies are not fast-evolving (see Section~\ref{sec:rq2-validation}); for other projects, developers need to tweak the configurations multiple times to find a sweet spot for their projects.
However, many respondents eventually find that Dependabot does not offer the options they want for noise reduction, such as update grouping and auto-merge.
We will investigate this in-depth in \textbf{RQ5}.

\begin{small}
\begin{result-rq}{Findings for RQ4:}
The majority of Dependabot configurations imply a proactive update strategy, but we observe multiple patterns of noise avoidance from configuration modifications, such as
increasing schedule intervals, lowering the maximum number of open PRs, and ignoring certain dependencies.
\end{result-rq}
\end{small}

\subsection{RQ5: Deprecations \& Desired Features}
\label{sec:rq5}


\subsubsection{Repository Analysis Methods}

To locate projects that may have deprecated Dependabot, we find projects with no \Code{dependabot.yml} in their current working trees, resulting in 235 projects.
For each of them, we identify the last commit that removes \Code{dependabot.yml}, inspect their commit messages, and identify any referenced issues/PRs following the GitHub convention.
If the \Code{dependabot.yml} removal turns out to be due to a project restructure or stop of maintenance, we consider it as a false positive and exclude it from further analysis.

For the remaining 206 projects, we analyze reasons for deprecation from commit messages and issue/PR text (i.e., titles, bodies, and comments). 
Since a large proportion of text in commit messages, issues, and PRs are irrelevant to Dependabot deprecation reasons, 
two authors read and re-read all text in the corpus, retaining only the relevant. 
They encode reasons from text and discuss them until reaching a consensus.
They do not conduct independent coding and measure inter-rater agreement because the corpus is very small (only 27 deprecations contain documented reasons).

For each of the confirmed deprecations, we check bot configuration files and commit/PR history to find possible migrations.
We consider a project as having migrated to another dependency management bot (or other automation approaches) if it meets any of the following criteria:
\begin{enumerate}[leftmargin=15pt]
    \item developers have specified the migration target in the commit message or issue/PR text;
    \item \Code{dependabot.yml} is deleted by another dependency management bot (e.g., Renovate Bot automatically deletes \Code{dependabot.yml} in its setup PR); 
    \item the project adopts another dependency management bot within 30 days before or after Dependabot deprecation.
\end{enumerate}

To obtain the developers' desired features for a dependency management bot, we ask two optional open-ended questions at the end of the survey (Table~\ref{tab:survey}).
The two questions are answered by 97 and 46 developers, respectively.
To identify recurring patterns from the answers, two authors of this paper (both with $>$6 years of software development experience and familiar with using Dependabot) conduct open coding~\cite{khandkar2009open} on the responses to generate an initial set of codes. 
They read and re-read all answers to familiarize themselves with and gain an initial understanding of them.
Then, one author assigns text in answers to some initial codes that reflects common features in dependency management bots and discusses with the other author to iteratively refine the codes until a consensus is reached.
They further conduct independent coding on the answers using the refined codes and exclude answers that do not reflect anything related to this RQ.
As each response may contain multiple codes, we use MASI distance~\cite{DBLP:conf/lrec/Passonneau06} to measure the distance between two raters' codes and Krippendorff’s alpha~\cite{krippendorff2018content} to measure inter-rater reliability.
The Krippendorff’s alpha we obtain is 0.865, which satisfies the recommended threshold of 0.8 and indicates a high reliability~\cite{krippendorff2018content}.

\subsubsection{Repository Analysis Results}
\label{sec:rq5-results}

\begin{figure}
  \centering
  \includegraphics[width=\linewidth]{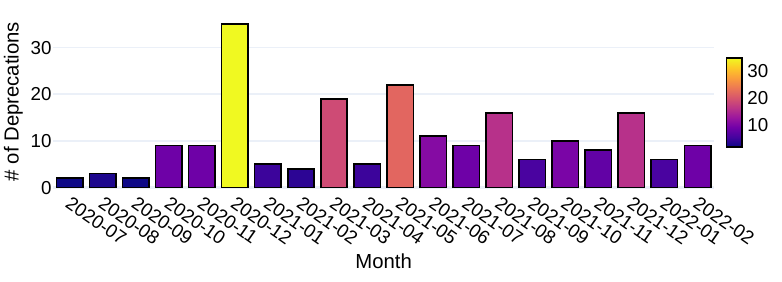}
\vspace{-5mm}
  \caption{Number of deprecations in each month.}
  \label{fig:deprecateDate}
\vspace{-3mm}
\end{figure}

We confirm 206 of the 235 candidates to be real-life Dependabot deprecations, which is substantial considering that our dataset only contains 1,823 projects. 
From Figure~\ref{fig:deprecateDate}, we can observe that Dependabot deprecations are evenly distributed over time in general with a few fluctuations, mostly coming from organization-wide deprecations.
For instance, the maximum value in December 2020 is caused by 26 Dependabot deprecations in \Code{octokit}, the official GitHub API client implementation.

We encode nine categories of reasons from the 27 deprecations that explicitly mentioned their reasons: 

\textbf{1) Notification Fatigue (9 Deprecations):}
Developers do recognize Dependabot's overwhelming notifications and PRs as the central issue in their experience with Dependabot.
As noted by one developer: \textit{I've been going mad with dependabot alerts which are annoying and pointless. I'd rather do manual upgrades than use this}~\cite{skytable/skytable/pull/134}.

\textbf{2) Lack of Grouped Update Support (7 Deprecations):}
By the Dependabot convention, each PR updates one dependency and one dependency only, which comes unhandy in two scenarios: 
a) related packages tend to follow similar release schedules, which triggers Dependabot to raise a PR storm on their updates~\cite{dependabot/dependabot-core/issues/1190};
b) in some cases, dependencies must be updated together to avoid breakages~\cite{dependabot/dependabot-core/issues/1296}.
The excessive notifications and additional manual work quickly frustrate developers.
For example: 
a) \textit{My hope was that we can better group dependency upgrades. With the default configuration, there is some grouping happening, but most dependencies would be upgraded individually}~\cite{giantswarm/happa/pull/2635};
b) \textit{Also, a lot of packages have to be updated together. Separate PRs for everything isn't very fun}~\cite{Fate-Grand-Automata/FGA/commit/8ecef22}.

\textbf{3) Package Manager Incompatibility (7 Deprecations):}
Developers may have compatibility issues after the introduction of a new package manager or a newer version of the package manager. 
In the seven cases we have found, five concern yarn v2, one concerns npm v7 (specifically lockfile v3), and one concerns pnpm.
To make matters worse, Dependabot may even have undesirable behaviors, e.g., messing around with yarn lockfiles~\cite{stoplightio/spectral/pull/1976}, when encountered with such incompatibilities. 
This contributes to developers' \textit{update suspicion}, as merging pull requests leads to possible breakages in dependency specification files.
At the time of writing, Dependabot still has no clear timeline on supporting pnpm~\cite{dependabot/dependabot-core/issues/1736} or yarn v2~\cite{dependabot/dependabot-core/issues/1297}.
For the unlucky part of Dependabot users, it means to revert~\cite{nitzano/gatsby-source-hashnode/issues/202}, to patch Dependabot PRs manually or automatically~\cite{replygirl/tc/issues/26}, or to migrate to an alternative, e.g., Renovate Bot~\cite{stoplightio/spectral/pull/1987}.

\textbf{4) Lack of Configurability (5 Deprecations):}
Dependabot is also deprecated due to developers' struggle to tailor a suitable configuration.
For example:
a) \textit{it appears that we're not able to configure Dependabot to only give us major/minor upgrades}~\cite{codalab/codalab-worksheets/pull/2916};
b) \textit{Dependabot would require too much configuration long-term -- too easy to forget to add a new package directory}~\cite{lyft/clutch/pull/126}.
Developers mention that other dependency management bots can provide more fine-grained configuration options such as update scope and schedule:
\textit{(Renovate Bot) has a load more options we could tweak too compared to Dependabot if we want to reduce the frequency further}~\cite{video-dev/hls.js/pull/3622}.

\textbf{5) Absence of Auto-Merge (3 Deprecations):}
Alfadel et al. \cite{DBLP:conf/msr/AlfadelCSM21} illustrate that auto-merge features are tightly associated with rapid PR merges. 
However, GitHub refused to offer this feature in Dependabot~\cite{dependabot/dependabot-core/issues/1973}, claiming that auto-merge allows malicious dependencies to propagate beyond the supervision of project maintainers. 
This may render Dependabot impractical, as claimed by a developer: \textit{(the absence of auto-merge) creates clutter and possibly high maintenance load}.

We notice a non-negligible proportion (8.17\%) of pull requests are merged by third-party auto-merge implementations (e.g., a CI workflow or a GitHub App). 
Unfortunately, they may become dysfunctional on public repositories after GitHub enforced a change on Dependabot PR triggered workflows~\cite{ahmadnassri/action-dependabot-auto-merge/issues/60}.
This turns out to be the last straw for several Dependabot deprecations.
As a developer states, they dropped Dependabot because \textit{latest changes enforced by GitHub prevent using the action in Dependabot's PR's context}.

\textbf{6) High CI Usage (3 Deprecations):}
Maintainers from 3 projects complain that Dependabot's substantial, auto-rebasing PRs have devoured their CI credits. 
In their words, Dependabot's CI usage is \textit{what killed us with Dependabot}, and \textit{a waste of money and carbon}.

Other reasons for Dependabot deprecation include: \textbf{7) Dependabot Bugs (2 Deprecations)}, \textbf{8) Unsatisfying of Branch Support (1 Deprecation)}, and \textbf{9) Inability to Modify Custom Files (1 Deprecation)}.




The deprecation of Dependabot does not necessarily mean developers' loss of faith in automating dependency updates. 
Actually, over two-thirds (68.4\%, 141/206) of the projects turn to another bot or set up custom CI workflows to support their dependency updates.
Among them, Renovate Bot (122) is the most popular migration target, followed by projen (15), npm-check-updates (2) and depfu (1).

\subsubsection{Triangulation from Survey}
\label{sec:rq5-validation}

Among the 131 surveyed developers, 14 (10.7\%) tell us they have deprecated Dependabot in their projects.
Most of the reasons they provide fall within our analysis and the frequency distribution is highly similar.
There are two exceptions: one deprecates because Dependabot frequently breaks code and one deprecates because their entire project has been stalled.
Developers also respond in our survey that they think automated dependency management is important and beneficial for their projects but the limitations of Dependabot causes them to do the deprecation. For example:
    \textit{Dependabot could be great, it just needs a few fixes here and there. It's unclear why Dependabot hasn't been polished.}
They also reply to us that Renovate Bot does provide some features that they need (e.g., grouped update PRs).

We identify nine major categories of developers' desired features (each corresponds to one code) from the answers provided by 84 respondents. 
The remaining categories are discarded as they are only supported by one answer (which thus may be occasional and not generalizable).
We will explain each category in the order of popularity.

\textbf{1) Group Update PRs (29 Respondents):}
This category refers to the feature of automatically grouping some dependency updates into one PR instead of opening one PR for each update.
It is most frequently mentioned and developers consider this feature as an important measure for making the handling of bot PRs less tedious, repetitive, and time-consuming. 
They want the bot to automatically identify dependencies that should be updated together and merge them into one PR update because \textit{many libraries (e.g., symfony, @typescript-eslint, babel) version all packages under a single version}.
They also want the bot to automatically find and merge ``safe'' updates into one PR while leaving ``unsafe'' updates as single PRs for more careful reviewing.

\textbf{2) Package Manager Support (20 Respondents):}
This category refers to the feature of supporting more package managers (and their corresponding ecosystems) or features for the bot to align with the conventions in the package manager/ecosystem.
Developers have expressed their desire for the bot to support Gradle, Flatter, Poetry, Anaconda, C++, yarn v2, Clojure, Cargo, CocoaPods, Swift Package Manager (in iOS), etc., indicating that dependency management bots, if well designed and implemented, can indeed benefit a wide range of developers and software development domains.
Dependabot does claim support for many package managers mentioned before but it still needs to be tailored and improved in, e.g., performance and update behaviors:
a) \textit{When I have 3 open Poetry updates I can merge one and then have to wait 15 minutes for the conflicts to be resolved.}
b) \textit{Perhaps for node.js projects the ability to update package.json in addition to package.lock, so the dependency update is made explicit.}

\textbf{3) Auto-Merge (19 Respondents):}
This category refers to the feature of automatically merging some update PRs into the repository if certain conditions are satisfied.
As mentioned in Section~\ref{sec:rq3-validation}, some developers believe as long as their projects have high-quality test suites, it will be trivial to review the update PR and they would prefer them to be merged automatically if the tests pass.

Despite the significant demand, this feature also seems to be especially controversial because doing this means offloading trust and giving bot autonomy.
Although GitHub considers it unacceptable due to security risks~\cite{dependabot/dependabot-core/issues/1973}, our survey clearly indicates that many still want to do this even if they are well aware of the risks.
They also think the responsibility of risk control, e.g., vetting new releases, should be given to some capable central authority, not them.
Here are three response examples:
a) \textit{While this might be somewhat dangerous, and should be configurable somehow, [auto-merge] is something that basically already happens when I merge such PRs.}
b) \textit{If I am merging with Dependabot like 60 deps a day - I don't know if some of the versions are not published by hackers who took over the repository account, so it would be great if there was some authority where humans actually check the changes and mark them secure.}
c) \textit{\textit{For me it'd be good if I could mute all notifications about Dependabot PRs except for when tests failed, indicating that I need to manually resolve some issues. Otherwise I'd be happy not to hear about it updating my deps.}}

\textbf{4) Display Release Notes (8 Respondents):}
This category refers to the feature of always showing some sort of release notes or change logs in update PRs to inform developers of the changes in an update.
Although Dependabot sometimes can provide release notes in PRs (Figure~\ref{fig:depbotpr}), it fails for 24.8\% of the PRs in our dataset.
One possible reason for this is that release notes are often missing or inaccessible in open source projects~\cite{2022ICPC-ReleaseNote}, which is also confirmed by one of our survey respondents: 
\textit{Most npm package updates feel unnecessary and the maintainers very often don't bother to write meaningful release notes...At the same time, I shouldn't expect maintainers to go through all of their dependencies' changelogs either, so perhaps the tool should find those release notes for me.}

\textbf{5) Avoid Unnecessary Updates (7 Respondents):}
This category refers to the feature of providing a default behavior and configuration options to avoid updates that most developers in an ecosystem perceived as unnecessary.
The most frequently mentioned feature is the ability to define separate update behaviors for development and production (or runtime) dependencies.
Many developers would avoid the automatic update of development dependencies because they perceive such updates as mostly noise and there is very little gain in keeping development dependencies up-to-date.
Other mentioned features include the ability to detect and avoid updates on bloated dependencies and to only provide updates for dependencies with real security vulnerabilities.

\textbf{6) Custom Update Action (5 Respondents):}
This category of features refers to the ability to define custom update behaviors (using, e.g., regular expressions) to update dependencies in unconventional dependency files.

\textbf{7) Configurability (5 Respondents):}
This category refers to the case of developers expressing that dependency management bots should be highly configurable, but does not provide any further information on the specific configuration options they want, e.g., \textit{more configuration options}.

\textbf{8) \Code{git} Support (4 Respondents):}
This category of features concerns the integration of dependency management bots with the version control system (in our case, \Code{git}).
The specific mentioned features include automatic rebase, merge conflict resolution, squashing, etc., all of which help ensure that bot PRs will not incur additional work on developers (e.g., manipulating \Code{git} branches and resolving conflicts).

\textbf{9) Breaking Change Impact Analysis (3 Respondents):}
This feature category refers to the ability to perform program analysis to identify breaking changes and their impact on client code, e.g.,
\textit{something like a list of parts of my codebase that might be impacted by the update would be useful. This could be based on a combination of changes listed in the release notes and an analysis of where the package is used in my code.} 

The developers' desired features align well with the reasons for Dependabot deprecation, indicating that feature availability can be an important driver for the migrations and competition between dependency management bots.

\begin{small}
\begin{result-rq}{Findings for RQ5:}
11.3\% of the studied projects have deprecated Dependabot due to notification fatigue, lack of grouped update support, package manager incompatibility, lack of configurability, absence of auto-merge, etc. 
68.4\% of them migrate to other ways of automation, among which the most common migration target is Renovate Bot (86.5\%).
We identify nine categories of developers' desired features that align well with the Dependabot deprecation reasons.
\end{result-rq}
\end{small}

\section{Discussion}
\label{sec:discussion}

\subsection{The State of Dependency Management Bots}

In a nutshell, our results indicate that Dependabot could be an effective solution for keeping dependency up-to-date (\textbf{RQ1}, \textbf{RQ2}), but often with significant noise and workloads (\textbf{RQ1}, \textbf{RQ4}, \textbf{RQ5}), many of which could not be mitigated by the features and configuration options offered by Dependabot (\textbf{RQ5}).
Apart from that, Dependabot's compatibility score solution is hardly a success in indicating the compatibility of a bot update PR (\textbf{RQ3}).
As of March 2023, Dependabot is still under active development by GitHub with the majority of effort in supporting more ecosystems (e.g., Docker, GitHub Actions) and adding features to reduce noise (e.g., automatically terminate Dependabot in inactive repositories), according to the GitHub change log~\cite{DependabotChangelog}.
Still, there is plenty of room for improvement to tackle the update suspicion and notification fatigue problem~\cite{DBLP:conf/kbse/MirhosseiniP17}.

Among other dependency management bots, Renovate Bot is an actively developed and popular alternative for Dependabot version update (\textbf{RQ5}), while Greenkeeper~\cite{Greenkeeper} has been deprecated, PyUp~\cite{PyUp} seems to be no longer under active development, and Synk Bot~\cite{SynkBot} mainly offers security-focused solutions.
As of March 2023, Renovate Bot provides more features and configuration options than Dependabot for fine-tuning notifications, including update grouping and auto-merge~\cite{RenovateDoc}; it also provides merge confidence badges with more information than Dependabot~\cite{RenovateMergeConfidence}.
However, it is still unclear whether the features and strategies taken by Renovate Bot are actually effective in practice, and we believe Renovate Bot could be an important study subject for future dependency management bot studies.

\subsection{What Should be the Key Characteristics of a Dependency Management Bot?}
\label{sec:implication}

In this section, we try to summarize the key characteristics of an ideal dependency management bot based on the results from our analysis and previous work.
We believe they can serve as general design guidelines for practitioners to design, implement, or improve dependency management bots (or other similar automation solutions).

\textbf{Configurability.} 
Wessel et al.~\cite{DBLP:journals/pacmhci/WesselWSG21} argue that noise is the central challenge in SE bot design and re-configuration should be the main countermeasure against noise.
For the case of Dependabot, we find that Dependabot also causes noise to developers by opening more PRs than developers can handle (\textbf{RQ2}), and developers can re-configure multiple times to reduce its noise (\textbf{RQ4}).
However, re-configuration is not always successful due to the lack of certain features in Dependabot, causing deprecations and migrations (\textbf{RQ5}).
Just as many other software development activities, it is also unlikely for a ``silver bullet'' to be present, as noted by one of our survey respondents, \textit{...there is no best practice in dependency management which is easy, fast and safe.}

Therefore, we argue that \textit{configurability}, i.e., offering the highest possible configuration flexibility for controlling its update behavior, should be one of the key characteristics of dependency management bots.
This helps the bot to minimize unnecessary update notifications and attempts so that developers are less interrupted.
Apart from the options already provided by Dependabot, our study indicates that the following configuration options should be present in dependency management bots:
\begin{enumerate}[leftmargin=15pt]
    \item \textit{Grouped Updates}: Dependency management bots should provide options to group multiple updates into one PR.
    Possible options include grouping all ``safe'' updates (e.g., not breaking the CI checks) and updates of closely related dependencies (e.g., different components from the same framework).
    \item \textit{Update Strategies}: Dependency management bots should allow developers to specify which dependency to update based on more conditions, such as whether the dependency is used in production, the severity of security vulnerabilities, whether the dependency is bloated, etc.
    \item \textit{Version Control System Integration}: Dependency management bots should allow developers to define how the bot should interact with the version control system, including which branch to monitor, how to manipulate branches and handle merge conflicts, etc.
\end{enumerate}

\textbf{Autonomy.}
According to the SE bot definition by Erlenhov et al.~\cite{DBLP:conf/sigsoft/ErlenhovN020}, the key characteristics of an ``Alex'' type of SE bot are its ability to autonomously handle (often simple) development tasks and its central design challenges include minimizing interruption and establishing trust with developers.
However, without the auto-merge feature, Dependabot is hardly autonomous and this lack of autonomy is disliked by developers (\textbf{RQ5}); in extreme cases, developers use Dependabot entirely as a notification tool but not as a bot (Section~\ref{sec:rq1-validation}).
This lack of autonomy is also causing a high level of interruption and workload to developers using Dependabot in their projects (\textbf{RQ5}).

We argue that \emph{autonomy}, i.e., the ability to perform dependency updates autonomously without human intervention under certain conditions, should be one of the key characteristics of dependency management bots.
This characteristic 
is only possible when the risks and consequences of dependency updates are highly transparent and developers \emph{know when to trust these updates}.
Within the context of GitHub, we believe the current dependency management bots should offer the configuration option to merge update PRs when the CI pipeline passes.
This option can be turned on for projects that have a well-configured CI pipeline with thorough static analysis, building, and testing stages, when the developers believe that their pipeline can effectively detect incompatibilities in dependency updates (Section~\ref{sec:rq3-validation}).

With respect to the security concern of \textit{auto-merge being used to quickly propagate a malicious package across the ecosystem}~\cite{dependabot/dependabot-core/issues/1973}, we argue that the responsibility of verifying new releases in terms of security should not be given to independent developers as they usually do not have the required time and expertise (\textbf{RQ5}).
Instead, package hosting platforms (e.g., npm, Maven, PyPI) should vet new package releases and quickly take down malicious releases to minimize their impact.
These practices are also advocated in the literature on software supply chain attacks~\cite{DBLP:conf/uss/ZimmermannSTP19}.

\textbf{Transparency.}
Multiple previous studies, both on SE bots and on other kinds of software bots, point to the importance of transparency in bot design.
For example, Erlenhov et al.~\cite{DBLP:conf/sigsoft/ErlenhovN020} shows that developers need to establish the trust that the bot can perform correct development tasks.
Similarly, Godulla et al.~\cite{godulla2021good} argue that transparency is vital for bots used in corporate communications.
In the context of code review bots, Peng and Ma~\cite{DBLP:journals/ccftpci/PengM19} find that contributors expect the bot to be transparent about why a certain code reviewer is recommended.
To reduce update suspicion~\cite{DBLP:conf/kbse/MirhosseiniP17} in dependency management bots, developers also need to know when to trust the bot to perform dependency updates.

We argue that \textit{transparency}, i.e., the ability to transparently demonstrate the risks and consequences of a dependency update, should be one of the key characteristics of dependency management bots.
However, the Dependabot compatibility score feature is hardly a success toward this direction and developers only trust their own test suites.
Beyond compatibility scores and project test suites, the following research directions may be helpful in enabling transparency in dependency management bots and establishing trust in the bot users:
\begin{enumerate}[leftmargin=15pt]
    \item \textit{Program Analysis}: One direction to achieve this is to leverage program analysis techniques.
    There have been significant research and practitioner effort on breaking change analysis~\cite{DBLP:conf/oopsla/LamDP20}, two of which have demonstrated the potential of using static analysis in assessing bot PR compatibility~\cite{DBLP:conf/sigsoft/FooCYAS18, DBLP:journals/jss/HejderupG22}.
    Still, given the extremely large scale of bot PRs~\cite{DBLP:conf/botse-ws/WyrichGHM21}, more research and engineering effort is needed to implement lightweight and scalable approaches to support each popular ecosystem.
    \item \textit{CI Log Analysis}: Another direction is to extend the idea of compatibility score with sophisticated techniques that learn more knowledge from CI checks. Since CI checks are scarce for many version pairs (\textbf{RQ3}), it will be interesting to explore techniques that transfer knowledge from other version pairs so that the matrix in Figure~\ref{fig:axios_compat} can be less sparse. 
    The massive CI checks available from Dependabot PRs would be a promising starting point.
    \item \textit{Release Note Generation}: Dependabot sometimes fails in locating and providing a release note for the updated dependency, and even if there is one, \textit{the maintainers very often don't bother to write meaningful release notes}, as noted by one respondent.
    This situation can be mitigated by applying approaches on software change summarization (e.g.,~\cite{DBLP:conf/scam/Cortes-CoyVAP14}) and release note generation (e.g.,~\cite{DBLP:journals/tse/MorenoBPOMC17}).
\end{enumerate}

\textbf{Self-Adaptability.}
The ability to adapt to the specific environment and its dynamics is considered as one of the key characteristics of a ``rational agent'' in artificial intelligence~\cite{poole1998computational, russell2010artificial}.
Dependency management bots can also be considered as autonomous agents working in the artificial environment of social coding platforms (e.g., GitHub).
However, our findings reveal that Dependabot often cannot operate in the ways expected by developers (\textbf{RQ5}) and re-configurations are common (\textbf{RQ4}).
Such failures (e.g., update actions, package manager incompatibility, \Code{git} branching) will lead to interruption and extra work for developers.

We argue that \emph{self-adaptability}, i.e., the ability to automatically identify and self-adapt to a sensible default configuration in a project's environment, should be one of the key characteristics of dependency management bots.
For GitHub projects, its environment can include its major programming languages, package managers \& ecosystems, the workflows used, the active timezone, developer preferences and recent activities, etc.
A dependency management bot should have the ability to automatically generate a configuration file based on such information, and recommend configuration changes when the environment has changed (e.g., developer responses to bot PRs become slower than usual).
This can be implemented by providing a semi-automatic recommender system for recommending an initial configuration to developers and prompting bot PRs for modifying their own configurations after bot adoption.

\vspace{-1.5mm}
\subsection{Comparison with Previous Work}
\vspace{-0.5mm}

Several previous studies have also made similar recommendations based on results from Greenkeeper or Dependabot~\cite{DBLP:conf/kbse/MirhosseiniP17, DBLP:conf/msr/AlfadelCSM21, rombaut2022there, cogo2022understanding}.
Studies on Greenkeeper~\cite{DBLP:conf/kbse/MirhosseiniP17, rombaut2022there} show that dependency management bot causes noise to developers and CI test results are unreliable, but they do not investigate the effectiveness of bot configurations as a countermeasure against noise.
Studies on Dependabot~\cite{DBLP:conf/msr/AlfadelCSM21, cogo2022understanding} either focuses on a different aspect (i.e., security updates~\cite{DBLP:conf/msr/AlfadelCSM21}) or provides specific recommendations on Dependabot features~\cite{cogo2022understanding}. 
Compared with the previous studies, the contributions of our study are: 1) a systematic investigation of the Dependabot version update service, and 2) a comprehensive four-dimension framework for dependency management bot design.

The implications of our study are also related to the larger literature of SE bots and dependency management.
With respect to the two fields, the contribution of our study is a unique lens of observation, i.e., Dependabot, that results in a set of tailored recommendations for dependency management bot design.
We have carefully discussed in Section~\ref{sec:implication} about how the implications of our study confirm, extend, or echo the implications from existing literature


\vspace{-0.5mm}
\subsection{Threats to Validity}
\label{sec:threats}

\subsubsection{Internal Validity}
\vspace{-0.25mm}

In \textbf{RQ1}, we have provided a holistic analysis of the impact of  Dependabot adoption without incorporating possible confounding factors (e.g., the types of dependencies and the characteristics of projects).
Consequently, it is difficult for our study to establish a firm answer on the effectiveness of adopting Dependabot and future work is needed to better quantify such impact among possible confounding factors.

Several approximations are used throughout our analysis.
In \textbf{RQ2}, we resort to identify security PRs ourselves which may introduce hard-to-confirm errors (only repository owners know whether their PRs are security-related).
The merge rate may not accurately reflect the extent to which Dependabot updates are accepted by developers as some projects may use different ways of accepting contributions.
To mitigate this threat, we focus on projects that have merged at least 10 Dependabot PRs with the intuition that these projects are unlikely to accept Dependabot PRs in other ways if they have already merged many of them.
In \textbf{RQ3}, Dependabot's compatibility scores may change over time and it is impossible to know the score at the time of PR creation.
In \textbf{RQ4}, Dependabot supports ecosystem specific matchers in dependency specifications, e.g., \Code{@angular/*}, which we do not consider when parsing configuration files.
However, we believe the noise introduced above should be minor and will not invalidate our findings or hinder the reproducibility of our data analysis.
Like other studies involving manual coding, our analysis of developer discussions and survey responses are vulnerable to author bias. 
To mitigate this, two authors double-check all results and validate findings with project commit/PR histories for \textbf{RQ5}; they further conduct inter-rater reliability analysis for \textbf{RQ5} when the dataset becomes larger. 
Finally, our own interpretation of the data (\textbf{RQ1} - \textbf{RQ5}) may also be biased towards our own judgment.
To mitigate this, we triangulate our key findings using a developer survey and derive implications based on both our analysis and developers' feedback.

\subsubsection{External Validity}
\vspace{-0.25mm}

Just like all case studies, generalizing our specific findings in each RQ to other dependency management bots and even to other projects that use Dependabot should be cautious.
Our dataset only contains popular and actively maintained GitHub projects, many of which are already taking proactive updating strategies. 
Therefore, our findings may not generalize to projects of a smaller scale or more reluctant to update dependencies. 
The survey responses are collected through convenience sampling which may introduce possible, yet unknown biases in terms of experience, age, gender, development role, etc., so the generalization of our survey results to a broad developer audience should be cautious.
The outcome of Dependabot usage may also not generalize to other dependency management bots due to their functionality and user base differences.
In \textbf{RQ1}, we only base our analysis on JavaScript/npm projects which may not generalize to other ecosystems with different norms, policies, and practices~\cite{DBLP:journals/tosem/BogartKHT21}; the comparison of dependency management bot usage in different ecosystems could be an important avenue for future work.
Despite these, we believe the implications we obtain for dependency management bot design should be general. 
Our proposed framework in Section~\ref{sec:implication} form a roadmap for dependency management bot designers.
Our methodology could be applied in future studies to compare the effectiveness of different bots. 




\vspace{-0.5mm}
\section{Conclusion}
\label{sec:conclusion}

We present an exploratory study on Dependabot version update service using repository mining and a survey, and we identify important limitations in the design of Dependabot. 
From our findings, we derive a four-dimension framework 
in the hope that it can help dependency management bot design and inspire more research work on related fields.

Several directions of future work arise from our study.
For example, investigating and comparing other dependency management bots, especially Renovate Bot, can help verify the generalizability of our proposed framework.
An empirical foundation on the factors affecting the effectiveness of bot adoption is also necessary.
It will be interesting to investigate the recommendation of bot configurations to developers, or to study how different approaches (e.g., program analysis, machine learning, release note generation) can help developers assess the compatibility of bot PRs.

\vspace{-0.5mm}
\section{Data Availability}

We provide a replication package at Figshare:

\begin{quote}
    \centering
    \url{https://figshare.com/s/78a92332e4843d64b984}
\end{quote}
The package can be used to replicate the results from repository mining.
To preserve the privacy of survey respondents, we choose not to disclose any raw data from the survey. 

\vspace{-0.5mm}
\section*{Acknowledgments}

This work is supported by the National Key R\&D Program of China Grant 2018YFB1004201 and the National Natural Science Foundation of China Grant 61825201.
We sincerely thank the developers who participated in our survey.

\bibliographystyle{IEEEtran}
\bibliography{refcomp}

\begin{IEEEbiography}[{
  \includegraphics[width=1in,height=1.25in,clip,keepaspectratio]{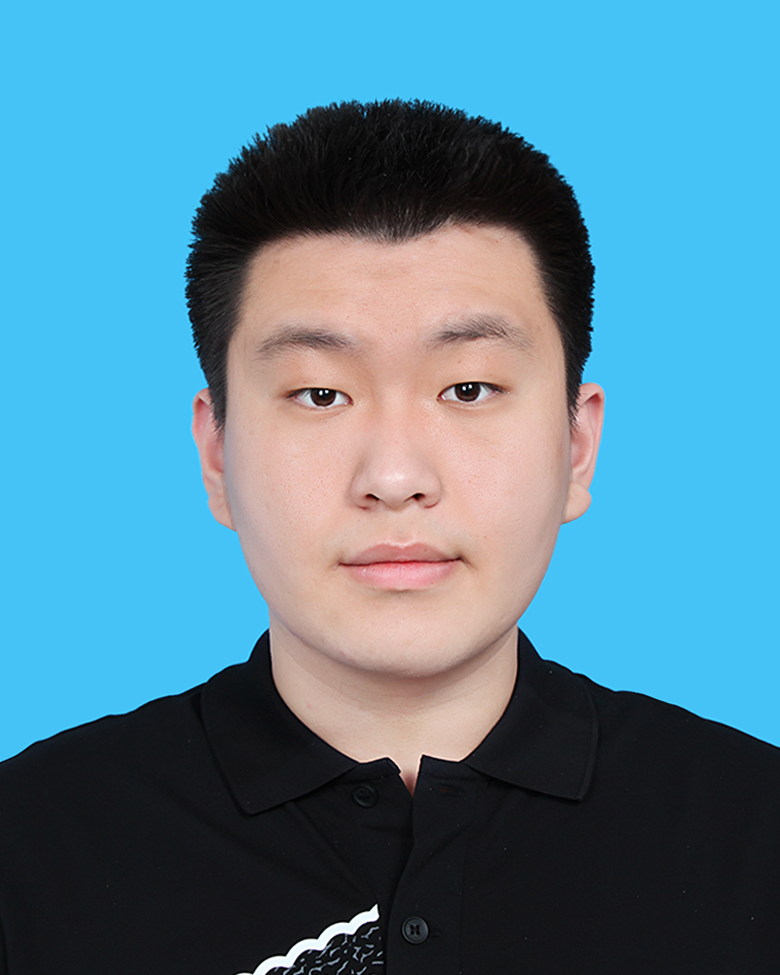}
}]{Runzhi He}
is currently an undergraduate student at the School of Electronics Engineering and Computer Science (EECS), Peking University.
His research mainly focuses on open source sustainability and software supply chain.
He can be contacted via \href{mailto:rzhe@pku.edu.cn}{rzhe@pku.edu.cn}
\end{IEEEbiography}

\begin{IEEEbiography}[{
  \includegraphics[width=1in,height=1.25in,clip,keepaspectratio]{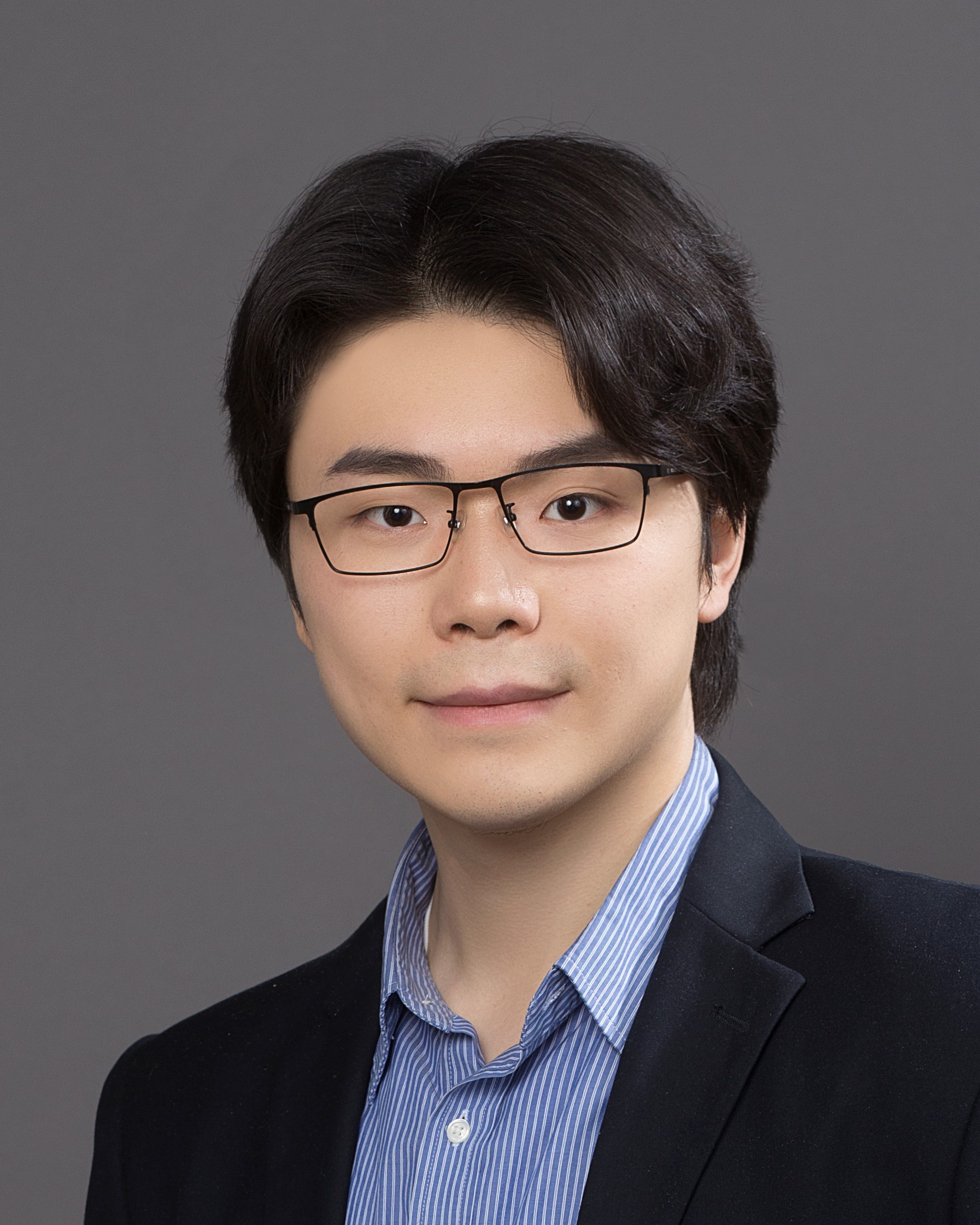}
}]{Hao He}
is currently a Ph.D. student at the School of Computer Science, Peking University. 
Before that, he received his B.S. degree in Computer Science from Peking
University in 2020.
His research addresses socio-technical sustainability problems in open source software communities, ecosystems, and supply chains.
More information can be found on his personal website \url{https://hehao98.github.io/} and he can be reached at \href{mailto:heh@pku.edu.cn}{heh@pku.edu.cn}.
\end{IEEEbiography}


\begin{IEEEbiography}[{
  \includegraphics[width=1in,height=1.25in,clip,keepaspectratio]{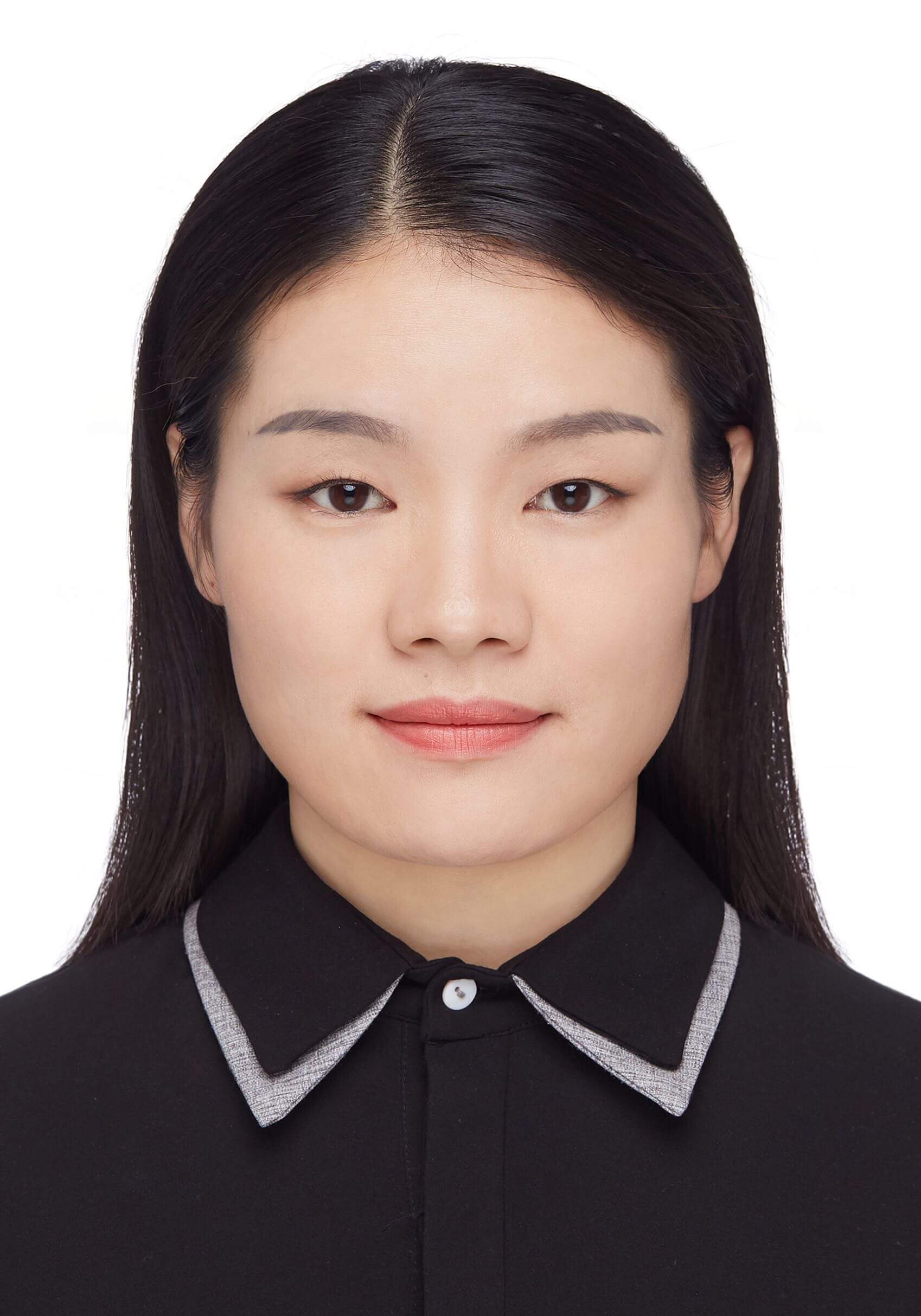}
}]{Yuxia Zhang} is currently an assistant professor at the School of Computer Science and Technology, Beijing Institute of Technology (BIT). She received her Ph.D. in 2020 from the School of Electronics Engineering and Computer Science (EECS), Peking University. Her research interests include mining software repositories and open-source software ecosystems, mainly focusing on commercial participation in open-source. She can be contacted at \href{mailto:yuxiazh@bit.edu.cn}{yuxiazh@bit.edu.cn}.
\end{IEEEbiography}

\begin{IEEEbiography}[{
  \includegraphics[width=1in,height=1.25in,clip,keepaspectratio]{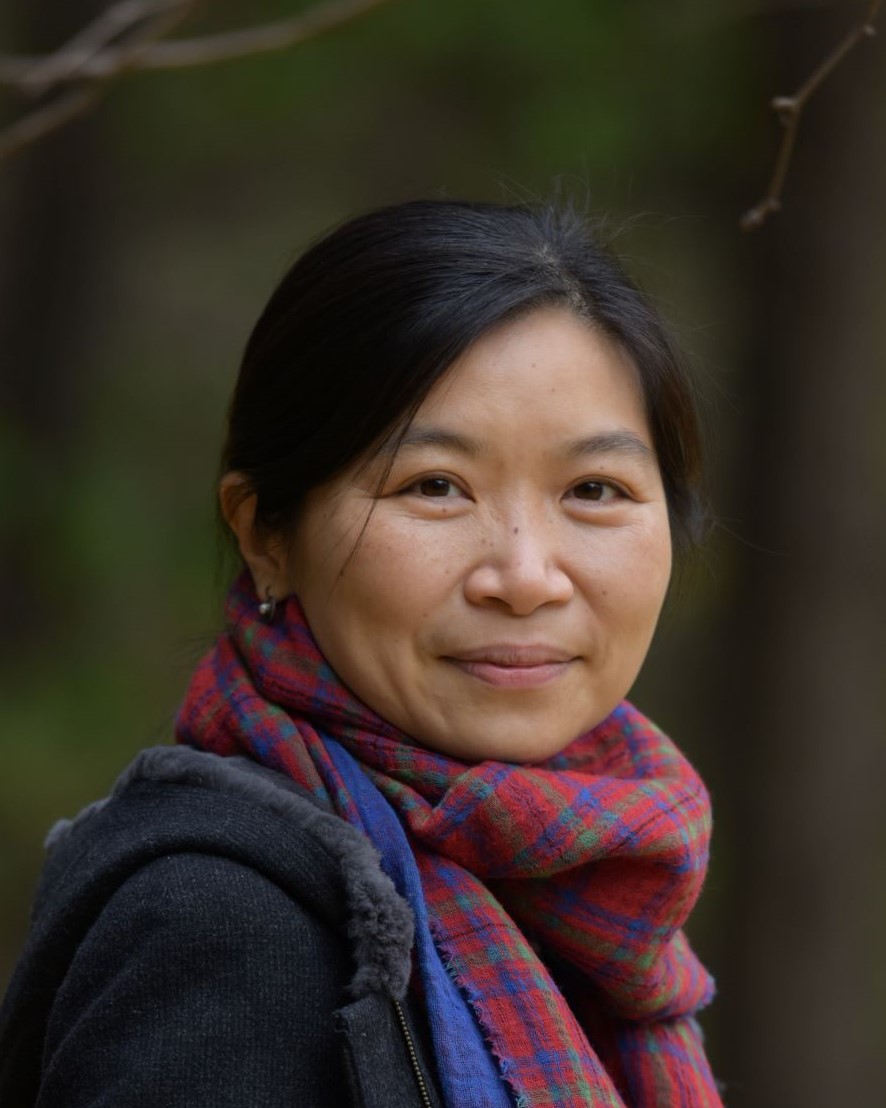}
}]{Minghui Zhou} received the BS, MS, and Ph.D. degrees in computer science from the National University of Defense Technology in 1995, 1999, and 2002, respectively. She is a professor in the School of Computer Science at Peking University. She is interested in software digital sociology, i.e., understanding the relationships among people, project culture, and software products through mining the repositories of software projects. She is a member of the ACM and IEEE. She can be reached at \href{mailto:zhmh@pku.edu.cn}{zhmh@pku.edu.cn}.
\end{IEEEbiography}

\end{document}